\documentclass[a4paper,11pt,reqno]{article}
\usepackage{a4wide}
\usepackage{amsmath,amsfonts,amssymb}
\usepackage[T1]{fontenc}
\usepackage[p,osf]{cochineal}
\usepackage[varqu,varl,var0]{inconsolata}
\usepackage[scale=.95,type1]{cabin}
\usepackage[vvarbb]{newtxmath}
\usepackage[cal=boondoxo]{mathalfa}
\usepackage[mathscr]{euscript}
\usepackage{bbold} 
\usepackage{pict2e}
\usepackage{lettrine} 
\usepackage{nicefrac}
\usepackage{setspace}
\usepackage{datetime}
\usepackage[nosort]{cite}
\usepackage{upgreek}

\usepackage{comment}
\usepackage[euler]{textgreek}
\usepackage[breaklinks=true]{hyperref}
\usepackage{cleveref}
\hypersetup{
			colorlinks=true,
			urlcolor=blue,
			citecolor=magenta,
			linkcolor=blue,
     }

\let\oldsqrt\sqrt
\def\sqrt{\mathpalette\DHLhksqrt}
\def\DHLhksqrt#1#2{%
\setbox0=\hbox{$#1\oldsqrt{#2\,}$}\dimen0=\ht0
\advance\dimen0-0.2\ht0
\setbox2=\hbox{\vrule height\ht0 depth -\dimen0}%
{\box0\lower0.4pt\box2}}

\newcommand{\RNum}[1]{\uppercase\expandafter{\romannumeral #1\relax}}

\author{
  \begin{minipage}{.97\linewidth}
    \vspace{1cm}
       \begin{center}
      \begin{small}
      \textbf{Nikolaos Athanasiou},$^1$
 \textbf{P. Marios Petropoulos},$^2$\\ 
      \textbf{Simon M. Schulz}$^3$
          and
      \textbf{Grigalius  Taujanskas}$^{4}$
              \end{small}
    \end{center}
    \vspace{0.5cm}
    \hspace{2.8cm}\begin{minipage}{.7\linewidth}
\begin{center}   
  {\it \begin{footnotesize}
\hbox{\kern-1.9cm\vbox{\begin{itemize}
 \item[$^1$]Department of Mathematics \& Applied Mathematics\\ 
       University of Crete\\
        Voutes Campus\\
        70013 Heraklion, Greece
      \end{itemize}
      \vskip0.23cm
     }
\kern-3.8cm\vbox{\vskip0.0cm
\begin{itemize}
 \item[$^2$] Centre de Physique Th\'eorique -- CPHT\\ 
        Ecole Polytechnique, CNRS\footnote{\emph{Centre National de la Recherche Scientifique}, Unit\'e Mixte de Recherche UMR 7644.}\\
        Institut Polytechnique de Paris\\
        91120 Palaiseau Cedex, France                \end{itemize}      
}}
\hbox{\kern-1.3cm\vbox{\begin{itemize}
 \item[$^3$] Scuola Normale Superiore\\ 
              Centro di Ricerca Matematica Ennio De Giorgi\\ 
              Piazza dei Cavalieri, 3\\
              56126 Pisa, Italy
      \end{itemize}
     }
\kern-4.3cm\vbox{\vskip0.0cm
\begin{itemize}
 \item[$^4$] Faculty of Mathematics\\ 
              University of Cambridge\\ 
              Wilberforce Road\\
              Cambridge CB3 0WA, UK
      \end{itemize}      
}}
     \end{footnotesize}}
\end{center}
    \end{minipage}
    \begin{center}
    \scriptsize
    \texttt{
    nikathan@act.edu,
   marios.petropoulos@polytechnique.edu,
   simon.schulz@sns.it, 
   taujanskas@dpmms.cam.ac.uk}
        \end{center}
     \end{minipage}
}

\title{\vspace{0.5cm}
 \boldmath \begin{Large}
    \textbf{\textsc{One-dimensional Carrollian fluids I: Carroll--Galilei duality}}
  \end{Large} \unboldmath
}

\date{}
\begin{document}

\begin{titlepage}
\maketitle
\thispagestyle{empty}

 \vspace{-12.3cm}
  \begin{flushright}
  CPHT-RR026.052024\\
  \end{flushright}
 \vspace{10.cm}

\begin{center}
\vspace{1.cm}
\textsc{Abstract}\\  
\vspace{0.6 cm}	
\begin{minipage}{1.0\linewidth}

Galilean and Carrollian algebras acting on two-dimensional Newton--Cartan and Carrollian manifolds are isomorphic. A consequence of this property is a duality correspondence between one-dimensional Galilean and Carrollian fluids. We describe the dynamics of these systems as they emerge from the relevant  limits of Lorentzian hydrodynamics, and explore the advertised duality relationship. This interchanges longitudinal and transverse directions with respect to the flow velocity, and permutes equilibrium and out-of-equilibrium observables, unveiling specific features of Carrollian physics. We investigate the action of local hydrodynamic-frame transformations in the Galilean and Carrollian configurations, i.e. dual Galilean and Carrollian local boosts, and comment on their potential breaking. Emphasis is laid on the additional geometric elements that are necessary to attain complete systems of hydrodynamic equations in Newton--Cartan and Carroll spacetimes. Our analysis is conducted in general Cartan frames as well as in more explicit coordinates, specifically suited to Galilean or Carrollian use. 
 
\end{minipage}
\end{center}

\vspace{6cm} 
\end{titlepage}

\onehalfspace

\begingroup
\hypersetup{linkcolor=black}
\tableofcontents
\endgroup
\noindent\rule{\textwidth}{0.6pt}

\section{Introduction}

The Galilei and Carroll algebras are distinct \.In\"on\"u--Wigner contractions of the Poincar\'e algebra, reached at infinite or zero velocity of light \cite{Levy, SenGupta,Henneaux:1979vn}. The corresponding groups act locally on the (co)tangent spaces of general Newton--Cartan or Carrollian manifolds  --- for a review of the field and further reading suggestions see \cite{NC, Duv, Duval:2014uoa, Duval:2014uva, Duval:2014lpa, Bekaert:2014bwa,Bekaert:2015xua, Morand:2018tke, Ciambelli:2019lap,Herfray:2021qmp}. In two spacetime dimensions, both Galilean and Carrollian algebras generate two translations and one boost; they are isomorphic.  This property has been exploited in the past in three-dimensional gravitational systems \cite{Bagchi:2010zz}. The purpose of the present note is to explore the consequences of this property for one-dimensional fluid dynamics on general Newton--Cartan and Carrollian spacetimes. For those, it  materializes as a duality which swaps time and space, and maps onto each other the geometric data as well as the fluid attributes such as the energy density, the stress, and the matter and heat currents following a very specific pattern typical of a strong/weak-coupling or a low/high-temperature relationship. Our goal here is twofold. First, to explore the above pattern and its potential consequences for defining Carrollian thermodynamics and in particular Carrollian entropy, temperature, and enthalpy, as mentioned in the concluding remarks of Sec. \ref{dualout}. And second, to explicitly derive full sets of hydrodynamic equations in both Galilean and Carrollian settings between which the duality is manifest, and to specialise at the end to flat Newton--Cartan and Carrollian geometries on which the rigorous analysis of the companion papers \cite{APST2,PST3} is carried out.

The dynamics of a Lorentzian fluid\footnote{This is also called a \emph{relativistic fluid}. Following  Jean-Marc L\'evy-Leblond, the latter is a misnomer because Galilean and Carrollian dynamics are also relativistic, although with a different relativity group.}  is encoded in the following equations (indices $\mu,\nu,\ldots$ take two values, $0$ and $1$, and refer to the coordinate basis, the coordinates being $x^0=ct$, $x^1=x$): 
\begin{equation}
\label{T-cons}
\nabla^\mu T_{\mu\nu}=0.
    \end{equation}
In this expression, $\text{T}=T_{\mu \nu}\text{d}x^\mu\text{d}x^\nu$  is the fluid (symmetric)  energy--momentum tensor. The pseudo-Riemannian spacetime is equipped with a metric $\text{d}s^2=g_{\mu\nu}\text{d}x^\mu\text{d}x^\nu$ and a Levi-Civita connection $\nabla$. 
A~relativistic fluid flows  along a time-like congruence $\text{u}$ with norm $\| \text{u} \|^2=-c^2$ ($c$ is the velocity of light). 

Following the conventions of Refs. \cite{Campoleoni:2018ltl, CMPR} for two-dimensional spacetime geometries, we introduce the Hodge-dual\footnote{Our conventions are:  $\ast u_\rho=u^\sigma \hat \eta_{\sigma\rho}$ with $ \hat \eta_{\sigma\rho}=\sqrt{\vert\det g\vert} \epsilon_{\sigma\rho}$ and $\epsilon_{01}=+1$. Hence $\hat\eta^{\mu\sigma}\hat\eta_{\sigma\nu}=\delta^\mu_\nu$.  The Hodge-dual of a vector is the Hodge-dual of the form with the index raised.}  $\ast\text{u}$, transverse  to $\text{u}$ and normalized as $\| \!\ast\!\text{u} \|^2=c^2$, and express the metric  in the (non-orthonormal) Cartan coframe $\{\text{u}, \ast \text{u} \}$ as\footnote{We systematically omit the tensor-product symbol $\otimes$ in this sort of expression.}
\begin{equation}
\text{d}s^2
=
\frac{1}{c^2}\left(-\text{u}^2+
 \ast \text{u}^2
\right).
\label{ds2gen}
\end{equation}
Furthermore
\begin{equation}
\label{chimag}
\text{q}=\chi \ast\! \text{u}, \quad \chi=-\frac{1}{c^2}\ast u^\mu T_{\mu\nu}u^\nu, \quad \varepsilon=\frac{1}{c^2}T_{\mu \nu} u^\mu u^\nu
\end{equation}
define the \emph{heat current}, the  \emph{heat density} and the \emph{energy density} of the fluid, whereas
the energy--momentum tensor reads:\footnote{The standard expression for the energy--momentum, combining longitudinal, mixed and transverse components with respect to the fluid velocity is $T_{\mu \nu}=(\varepsilon+p) \frac{u_\mu  u_\nu}{c^2} +p  g_{\mu\nu}+   \tau_{\mu \nu}+ \frac{u_\mu  q_\nu}{c^2}   +\frac{u_\nu  q_\mu}{c^2}$. \label{Tmunu}}
\begin{equation}
\text{T}=\frac{1}{2c^2}\left(\left(\varepsilon+\chi\right)\left(\text{u}+\ast \text{u}\right)^2+\left(\varepsilon-\chi\right)\left(\text{u}-\ast \text{u}\right)^2\right)
+\frac{1}{c^2}(p-\varepsilon+\tau) \ast\! \text{u}^2
\label{Tgen}
\end{equation}
with trace  $T^\mu_{\hphantom{\mu}\mu}=p-\varepsilon+\tau$.
In this  expression,  $\tau$ is the \emph{viscous stress scalar}, the unique component of the viscous stress tensor 
\begin{equation}\label{viscstr}
 \tau_{\mu \nu}=\dfrac{\tau}{c^2}\ast\! u_{\mu} \ast\!u_{\nu}
\end{equation}
($h_{\mu\nu} = \nicefrac{1}{c^2} \ast \! u_{\mu} \ast \! u_{\nu}$ is the projector onto the space transverse to the velocity field). The separation between $p$ and $\tau$ is arbitrary from the geometric perspective since they are both fully transverse, $p+\tau$ being the total stress --- as opposed to the heat current $\text{q}=\chi \ast\! \text{u}$, which enters the energy--momentum tensor as a mixed component $\text{q} \text{u}=\chi \ast\! \text{u}\,\text{u}$, while the energy density is the longitudinal component. If the system is free and at global equilibrium, $\tau$ vanishes and the stress is the thermodynamic pressure $p$. More generally, the viscous stress scalar $\tau$ is expressed through a \emph{constitutive relation}, i.e. as an expansion in temperature and velocity gradients, and this distinguishes it from  $p$. The same holds for the heat density $\chi$. The fluid equations \eqref{T-cons} are usually supplemented with a local equation of state $\varepsilon=\varepsilon(\upsigma, p)$, where $\upsigma$ is the relativistic entropy (not to be confused with the stress $\sigma$ introduced later in \eqref{varlimga} and \eqref{varlimcar} for Galilean and Carrollian fluids respectively).

In two spacetime dimensions, the only non-vanishing first-derivative  tensors of the velocity are the expansion scalars
$\Theta=\nabla_\mu  u^\mu
$  
and $\Theta^{\ast}=\nabla_\mu  \ast \! u^\mu 
$,
equivalently defined as the velocity-form exterior differentials\footnote{Note that $\Theta^*$ is a scalar. For any vector $\text{v}$ and a function $h$, $\text{v}(h)=v^\mu\partial_\mu h$. We remind that 
$\text{d}h= \frac{1}{c^2}\left(\ast \text{u}(h)\ast \!\text{u}- \text{u}(h)\text{u} \right)$,
$\ast\text{d}h= \frac{1}{c^2}\left(\ast \text{u}(h)\text{u}-\text{u}(h) \ast \! \text{u} \right)$, 
and quote also $\ast\!\left(\text{u}\wedge \ast\text{u}\right)=c^2$.} 
\begin{equation}
\text{d}\ast\!\text{u} =\frac{\Theta}{c^2}\ast \! \text{u} \wedge\text{u} 
\quad \text{and} \quad
\text{d}\text{u} =\frac{\Theta^\ast}{c^2}\ast \! \text{u} \wedge \text{u},
 \label{def21eq}
\end{equation}
or in the Lie derivatives of the vectors (we do not make any typographic distinction between the 1-form $\text{u}$ and the vector $\text u$)
\begin{equation}
\left[\text{u},\ast\text{u} 
\right]=
\Theta^\ast 
\text{u} 
-\Theta
\ast\!\text{u}.
 \label{def21eqv}
\end{equation}
All information about the Levi-Civita connection in the frame $\{\text{u}, \ast \text{u} \}$ is encoded in $\Theta$ and $\Theta^\ast$. In particular, the acceleration is 
$
a_\mu =u^\nu \nabla_\nu u_\mu  =\Theta^{\ast}
\ast \!u_\mu
$. 
Projecting Eqs.  \eqref{T-cons} onto $\text{u}$ and $\ast\text{u}$, we obtain the longitudinal (energy) and transverse (momentum) relativistic fluid equations
\begin{equation}
\label{T-cons-el-mag-nc-force-equations} 
\mathcal{L}\equiv -u^\nu \nabla^\mu T_{\mu\nu}=
0,\quad
c\mathcal{T}\equiv \ast u^\nu \nabla^\mu T_{\mu\nu}=
0
\end{equation} 
with 
\begin{equation}
\label{T-cons-el-mag-nc-force} 
 \begin{cases}
\mathcal{L}=
\text{u}(\varepsilon)+  \Theta(p+\tau+\varepsilon)
+ \ast\text{u}(\chi)+ 2\Theta^\ast\chi 
,\\
c\mathcal{T}=\text{u}(\chi)+ 2\Theta \chi
+
\ast\text{u}(p+\tau)+ \Theta^\ast (p+\tau+\varepsilon
).
\end{cases}
\end{equation} 

It has been  known since Eckart  (see Refs.  \cite{Eckart, Landau}, and  for a recent review \cite{Kovtun:2012rj})  that one can perform local Lorentz boosts on the velocity congruence, while keeping unaltered the entropy current and the energy--momentum tensor. Physically, this reflects the indistinguishability of relativistic energy and mass flows, making the velocity field non-measurable (except for the case of ideal fluids). For non-relativistic, i.e. Galilean fluids, the velocity is measurable however, suggesting that the hydrodynamic-frame invariance is broken in the infinite-$c$ limit. This issue was settled  in \cite{BigFluid}, where it was shown that the breaking  occurs in the absence of  subleading terms (of order $\nicefrac{1}{c^2}$) in the relativistic matter current. When such terms are available, the invariance is maintained at the expense of dealing with sinks or springs of matter, which enter among others the continuity equation. This phenomenon fits the advertised duality relating Galilean with Carrollian fluids, and is therefore part of our agenda. 

In two spacetime dimensions the Lorentz group is one-dimensional. It is parameterized by the rapidity $\psi^\ast=\psi^\ast(t,x)$ and acts on  $\{\text{u}, \ast \text{u} \}$ as
\begin{equation}
\label{finiteLor}
\begin{pmatrix}
\text{u}^\prime\\
\ast \text{u}^\prime\
\end{pmatrix}
=
\begin{pmatrix}
\cosh \psi^\ast&\sinh \psi^\ast \\
\sinh \psi^\ast&\cosh \psi^\ast
\end{pmatrix}
\begin{pmatrix}
\text{u}\\
\ast \text{u}
\end{pmatrix},
\end{equation} 
or infinitesimally\footnote{Accordingly, $\delta_{\text{L}} \Theta=\psi^\ast\Theta^\ast+\ast \text{u}\left(\psi^\ast\right)
$ and $\delta_{\text{L}} \Theta^\ast=
\psi^\ast \Theta+\text{u}\left(\psi^\ast\right)$.\label{deltatheta}}
\begin{equation}
\label{locLor}
\delta_{\text{L}} \text{u}=  \psi^\ast \ast \! \text{u},
\quad
\delta_{\text{L}} \ast\! \text{u}
=\psi^\ast 
\text{u}.
\end{equation}
The invariance of the energy--momentum tensor implies thus\footnote{For a finite transformation
$
\begin{pmatrix}
\varepsilon'&\chi'\\
\chi'&p'+\tau'
\end{pmatrix}
=
\begin{pmatrix}
\cosh \psi^\ast&-\sinh \psi^\ast \\
-\sinh \psi^\ast&\cosh \psi^\ast
\end{pmatrix}
\begin{pmatrix}
\varepsilon&\chi\\
\chi&p+\tau
\end{pmatrix}
\begin{pmatrix}
\cosh \psi^\ast&-\sinh \psi^\ast \\
-\sinh \psi^\ast&\cosh \psi^\ast
\end{pmatrix}
$
.
}
\begin{equation}
\label{locLoremt}
\delta_{\text{L}} \varepsilon= \delta_{\text{L}} (p+\tau)
=-2 \psi^\ast \chi,
\quad
\delta_{\text{L}} \chi
=- \psi^\ast (\varepsilon +p+\tau).
\end{equation}
Equations \eqref{T-cons} are invariant under these transformations because the components $T_{\mu\nu}$ are invariant as is the connection. 
However, \eqref{T-cons-el-mag-nc-force} are not due to the projections onto the Cartan basis $\{\text{u}, \ast \text{u} \}$, which transforms, and we find:
 \begin{equation}
\label{locLorequations}
\delta_{\text{L}}\mathcal{L}=  - c \psi^\ast \mathcal{T},
\quad
\delta_{\text{L}}\mathcal{T}
=-\frac{\psi^\ast }{c}
\mathcal{L}.
\end{equation}
Nonetheless, the hydrodynamic equations \eqref{T-cons-el-mag-nc-force-equations} are of course invariant on-shell.

The Cartan frame  $\{\text{u}, \ast \text{u} \}$ is rather abstract and in practice we  need a  parameterization of $\text{u}$ and $\ast\text{u}$ in terms of four arbitrary functions,\footnote{The function $\gamma$ here is not to be confused with the constant constitutive exponent $\gamma$, which plays a role in the companion papers \cite{APST2,PST3}.} $\Gamma$, $\Delta_x$, $v^x$ and $\gamma$, of two coordinates $\{t,x\}$. 
The expressions for  the vectors are
\begin{equation}
\label{genv}
\text{u} =\gamma\left(\partial_t+v^x\partial_x\right),
\quad \ast \text{u} = \frac{c}{\Gamma} \left(\partial_x
+\Delta_x \gamma \left(\partial_t+v^x\partial_x\right)
\right),
\end{equation}
where $v^x$ is the physical velocity of the fluid along $x$ and  $\gamma$ is its Lorentz factor, whereas their dual forms read:\footnote{Using Eqs. \eqref{def21eq} we find: $
\Theta=\frac{\gamma}{\Gamma}\left(\partial_t\Gamma
+\partial_x\left(\Gamma v^x\right) 
\right)
$ and $
\Theta^\ast=\frac{c}{\Gamma}\left(-\partial_x\ln\gamma+\gamma\left(\partial_t \Delta_x+\partial_x\left(\Delta_x v^x\right)\right)\right)
$.\label{relexp}
} 
\begin{equation}
\label{genf}
\text{u} = c^2 \left(-\frac{\text{d}t}{\gamma}+\Delta_x\left(\text{d}x-v^x\text{d}t\right)\right),
\quad \ast \text{u} = c\Gamma \left(\text{d}x-v^x\text{d}t\right).
\end{equation}
The pseudo-Riemannian metric \eqref{ds2gen} and its inverse, the cometric, then assume the form:
\begin{eqnarray}
\label{genmetr}
\text{d}s^2
&=&-c^2 \left(
\frac{\text{d}t}{\gamma}-\Delta_x\left(\text{d}x-v^x\text{d}t\right)
\right)^2
+\Gamma^2
\left(\text{d}x-v^x\text{d}t\right)^2,\\
\label{gencometr}
\partial_s^2
&=&-\frac{\gamma^2}{c^2}\left(\partial_t+v^x\partial_x\right)^2
+\frac{1}{\Gamma^2}\left(\partial_x
+\Delta_x \gamma \left(\partial_t+v^x\partial_x\right)
\right)^2.
\end{eqnarray}

In Secs. \ref{galfl} and \ref{carfl}, we present the Galilean and Carrollian hydrodynamic equations as they emerge from their common relativistic ascendant. 
The duality relating these two dynamical systems is discussed in Sec. \ref{dualout}, and should be considered the central result of our work. Additionally, as already mentioned,
the hydrodynamic-frame invariance originating from the Lorentzian system is thoroughly analyzed and proven to hold both for Galilean and Carrollian descendants, in line with the duality relating them. On the Galilean side, this property is rooted in the possibility of creating or destroying matter. This phenomenon is normally unwanted. Avoiding it requires setting to zero the loss/gain current, which
spoils the Galilean hydrodynamic-frame invariance and promotes the Galilean velocity field to a measurable physical observable.  On the Carrollian  side, a dual breaking is reached by demanding the vanishing of the Carrollian heat current. As highlighted earlier,  
this features the specific  duality property revealed in this work, which maps the geometric or kinematic data onto the energy density, the stress, and the matter and heat currents.

The number of Galilean or Carrollian hydrodynamic equations yielded by expanding \eqref{T-cons} is four rather than the expected two --- one in time for the energy, and one in space for the momentum, or Euler. This is achieved following \cite{BigFluid,CMPPS1} by keeping the appropriate orders in the $c^2$-expansion of the relativistic energy-momentum tensor, and amounts to the existence of new Galilean or Carrollian momenta. If these were to be interpreted as variations of a microscopic action with respect to background fields, one would require more than the standard geometric fields. Although we do not delve into this investigation, the extra fields are naturally incorporated in our analysis in the form of subleading terms in the expanded Lorentzian frames (see \eqref{uvecexp}, \eqref{ufexp} for Galilei, and \eqref{uvecexpc}, \eqref{ufexpc} for Carroll). This extra geometric structure on Newton--Cartan or Carroll manifolds is new and deserves to be emphasized. It is
also worth stressing that local boost invariance of a microscopic action is broken if certain specific momenta are non-vanishing ($P_x$ defined in  \eqref{galMZ} for the Galilean and the dual $\Pi^x$ appearing in \eqref{carEPR} for the Carrollian instance --- see e.g.\cite{duval1978,dutch}). We emphasize that the hydrodynamic equations, however, are boost-invariant (irrespective of the dimension), a subtlety which appears to be frequently overlooked in the literature.

\section{Galilean fluids}\label{galfl}

\subsection*{Newton--Cartan geometries and non-relativistic fluids}

Galilean fluids are ordinary non-relativistic fluids, flowing however on more general spacetimes than the direct product of real time and Euclidean space. 
These are Newton--Cartan geometries 
\cite{NC, Duv, Duval:2014uoa,Bekaert:2014bwa, Bekaert:2015xua,Morand:2018tke}. In a nutshell, a Newton--Cartan geometry is a spacetime equipped with a \emph{degenerate cometric} $\partial_\ell^2$. The kernel of this cometric is the \emph{clock form} $\uptau$, which has a dual\footnote{Duality refers here to the linear action of forms on vectors: $\langle\uptau, \upupsilon\rangle=\iota_{\upupsilon}\uptau=1$ (not to be confused with the scalar product, which is precisely degenerate here).} vector field, the \emph{field of observers}  $\upupsilon$. Newton--Cartan spacetimes arise as infinite-$c$ limits of Lorentzian geometries, as one observes in Eq. \eqref{gencometr}. 

Dynamics for continuous media can be obtained from first principles directly in the Newton--Cartan framework \cite{duval1978,Geracie:2015xfa, Festu,Armas:2019gnb}. Alternatively, and perhaps more physically, following the pattern originally presented in \cite{Landau}, one can reach this dynamics in the non-relativistic limit, i.e. at infinite $c$, from the relativistic fluid equations. This is the point of view we shall pursue here because it   enables us to recover elegantly the full set, including auxiliary equations, in the spirit of Refs.  \cite{BigFluid, CMPPS1}, but working in a frame suited to our two-dimensional spacetime purposes.

To that end we must keep track of the subleading orders appearing in the Galilean limit of the Lorentzian geometric data, which are the forms and vectors $\text{u}$ and $\ast \text{u}$. Expanding Eqs. \eqref{genv} and \eqref{genf} in powers of $c^{-1}$, at large $c$ the vectors behave as 
\begin{equation}
\label{uvecexp}
\text{u} = \upupsilon +\frac{\upupsilon^{(2)}}{c^2}+\cdots, \quad \ast\text{u} = c\left(\ast\upupsilon +\frac{\ast\upupsilon^{(2)}}{c^2}+\cdots\right),
\end{equation}
for some vectors $\upupsilon$, $\upupsilon^{(2)}$, $\ast\upupsilon$ and $\ast\upupsilon^{(2)}$, whereas the forms give
\begin{equation}
\label{ufexp}
\text{u} = c^2 \upmu +\upmu_{(2)}+\cdots, \quad \ast\text{u} = c\left(\ast\upmu +\frac{\ast\upmu_{(2)}}{c^2}+\cdots\right)
\end{equation}
for some forms $\upmu$, $\upmu_{(2)}$, $\ast\upmu$ and $\ast\upmu_{(2)}$.
The duality relations $\langle\text{u}, \text{u}\rangle=-c^2$,  $\langle\text{u}, \ast\text{u}\rangle=\langle\ast\text{u}, \text{u}\rangle=0$ and $\langle\ast\text{u}, \ast\text{u}\rangle=c^2$  translate into\footnote{Together with $\left\langle\upmu, \upupsilon^{(2)}\right \rangle+\left\langle\upmu_{(2)}, \upupsilon \right\rangle=\left\langle\ast\upmu, \upupsilon^{(2)}\right \rangle+\left\langle\ast\upmu_{(2)}, \upupsilon \right\rangle=\left\langle\upmu, \ast\upupsilon^{(2)}\right \rangle+\left\langle\upmu_{(2)}, \ast\upupsilon \right\rangle=
 \left\langle\ast\upmu, \ast\upupsilon^{(2)}\right \rangle+\left\langle\ast\upmu_{(2)}, \ast\upupsilon \right\rangle=0$.\label{dual2}} 
\begin{equation}
\label{dualgal}
\langle\upmu, \upupsilon \rangle=-1, \quad \langle\upmu, \ast\upupsilon\rangle=0, \quad \langle\ast\upmu, \upupsilon\rangle=0, \quad \langle\ast\upmu, \ast\upupsilon\rangle=1,
\end{equation}
and we find the following behaviour for the cometric:
\begin{equation}
\partial_\ell^2=\lim_{c\to \infty} \partial_s^2=\ast \upupsilon^2.
\end{equation}
This bilinear form is degenerate with clock-form kernel $\uptau=-\upmu$ and field of observers $\upupsilon$.  We have thus reached a Newton--Cartan spacetime with tangent space spanned by $\{\upupsilon, \ast \upupsilon \}$, and cotangent space by $\{\upmu, \ast \upmu \}$ with $\ast \upmu $ called the \emph{ruler form}.

Several remarks are in order here. First, one should stress that there is no Hodge duality in the Newton--Cartan geometry because the cometric is degenerate.\footnote{In higher dimensions, we can define a spatial Hodge duality though, see e.g. \cite{BigFluid}.}  The stars are just notational remnants which  provide a convenient discrimination of the two directions. Next, observe that the vectors and forms $\left\{\upupsilon^{(2)}, \ast \upupsilon^{(2)} \right\}$ and $\left\{\upmu_{(2)}, \ast \upmu_{(2)} \right\}$ are not part of the definition of the Newton--Cartan geometry. Instead, they provide extra information which is necessary for achieving the complete set of hydrodynamic equations. Higher-order terms in Eqs. \eqref{uvecexp} and \eqref{ufexp}, i.e. terms of order $c^{-m}$, $m> 2$, do not contribute to the dynamics because they turn out to affect the terms of order $c^{-n}$, $n\geq 2$, in expressions \eqref{eqlimga}, which vanish at infinite $c$.  Contributions of order $c^2$ and $c^4$ in  \eqref{uvecexp} and \eqref{ufexp}, however, would bring extra terms of order $c^4$ in \eqref{eqlimga}, hence two more Galilean  equations. This would not be standard hydrodynamics, but might have some interest in its own right (see e.g. \cite{BigFluid} for a related discussion).

For the expansion scalars we obtain similarly:
\begin{equation}
\Theta=\theta+\frac{\theta^{(2)}}{c^2}+\cdots, \quad \Theta^\ast = c\left(\theta^\ast+\frac{\theta^{\ast(2)}}{c^2}+\cdots\right).
\end{equation}
The various functions emerging in these expansions appear in the exterior differentials as in \eqref{def21eq},
\begin{equation}
\text{d}\ast\!\upmu =\theta\ast \! \upmu \wedge\upmu ,
 \quad
\text{d}\upmu =\theta^\ast\ast \! \upmu \wedge \upmu,
 \label{def21eqgal}
\end{equation}
and
\begin{eqnarray}
&&\text{d}\ast\!\upmu_{(2)} =\theta^{(2)} \ast\! \upmu \wedge\upmu 
+ \theta\left( \ast\upmu \wedge\upmu_{(2)} +\ast \upmu_{(2)}  \wedge\upmu \right), 
 \label{def21eqgal2} \\
&&\text{d}\upmu_{(2)} =\theta^{\ast(2)} \ast\! \upmu \wedge\upmu 
+ \theta^\ast\left( \ast\upmu \wedge\upmu_{(2)} +\ast \upmu_{(2)}  \wedge\upmu \right),
 \label{def21eqgal2ast}
\end{eqnarray}
or alternatively in the Lie brackets of the vectors $\left[\upupsilon,\ast\upupsilon 
\right]$, $\left[\upupsilon,\ast\upupsilon^{(2)} 
\right]$ and $\left[\upupsilon^{(2)} ,\ast\upupsilon 
\right]$ (see Eq. \eqref{def21eqv}). 

In order to take the Galilean limit of the relativistic equations $\mathcal{L}=0$ and $\mathcal{T}=0$, we must prescribe the large-$c$ behaviour of the physical variables: the energy density, the heat current and the stress tensor. Following \cite{Landau, BigFluid, CMPPS1}, we assume that the pressure $p$ remains unaltered in the infinite-$c$  limit and keep the same notation. For the others we set
\begin{equation}
\varepsilon = \varrho c^2 + \tilde e + \text{O}\left(\frac{1}{c^2}\right),
 \quad 
 \tau=\sigma + \text{O}\left(\frac{1}{c^2}\right),
 \quad
\chi=c \tilde\zeta + \frac{\zeta}{c}+ \text{O}\left(\frac{1}{c^3}\right).
 \label{varlimga}
\end{equation}
In these expressions, $\varrho$ is the rest-energy contribution to be interpreted as the non-relativistic mass density, $\tilde e$ is a combination of internal and kinetic energies, whereas $\sigma$ and $\zeta$ are the non-relativistic stress and the non-relativistic heat current. The $\tilde\zeta$ contribution is 
unusual because it is a dominant term in the relativistic heat current. Such a term is normally absent because it turns out to mimic creation or loss of mass--energy and alters the continuity equation. Without it, however, relativistic hydrodynamic-frame invariance does not survive in the Galilean limit, and the advertised duality with the Carrollian equations fails. 

The large-$c$ expansion of $\mathcal{L}$ and $\mathcal{T}$ displayed in Eqs. \eqref{T-cons-el-mag-nc-force} is finally achieved as 
\begin{equation}
\mathcal{L}= c^2 \mathcal{C}+\mathcal{E} + \text{O}\left(\nicefrac{1}{c^2}\right)
,
\quad
\mathcal{T}= c^2 \mathcal{N}+\mathcal{M} + \text{O}\left(\nicefrac{1}{c^2}\right)
\label{eqlimga}
\end{equation}
with 
\begin{eqnarray}
\mathcal{E} &=&\upupsilon(\tilde e)+ \theta(p+\sigma+\tilde e)
+ \ast\upupsilon(\zeta)+ 2\theta^\ast \zeta
\nonumber\\
&&+\ \upupsilon^{(2)}(\varrho)+ \theta^{(2)} \varrho
+ \ast\upupsilon^{(2)}\left( \tilde\zeta\right)+ 2\theta^{\ast(2)} \tilde\zeta,
\label{galE}
\\
\label{galM}
\mathcal{M} &=&\ast\upupsilon(p+\sigma)+ \theta^\ast(p+\sigma+\tilde e) +\upupsilon\left( \tilde\zeta\right)+ 2\theta \tilde\zeta+\theta^{\ast(2)}  \varrho,
\\
\label{galC}
\mathcal{C} &=& \upupsilon (\varrho)+ \theta \varrho
+ \ast\upupsilon\left( \tilde\zeta\right)+ 2\theta^{\ast} \tilde\zeta,
\\
\label{galN}
\mathcal{N} &=&\theta^{\ast}  \varrho.
\end{eqnarray}
By demanding that $\mathcal{L}$ and $\mathcal{T}$ vanish for all values of $c$ and taking the limit $c\to \infty$, we derive four distinct equations corresponding to the vanishing of $\mathcal{E}$, $\mathcal{M}$, $\mathcal{C}$, and $\mathcal{N}$. Although the present formalism is somewhat abstract, an interpretation of these equations is possible owing to the fact that $ \upupsilon $ and  $ \ast \upupsilon $ are, effectively, the time and space derivatives respectively. The interpretations are as follows.
\begin{itemize}
\item The energy equation is $\mathcal{E} =0$.
\item The momentum (Euler) equation is $\mathcal{M} =0$.
\item The continuity equation is $\mathcal{C} =0$.
\item The equation $\mathcal{N} =0$ is more exotic. It is satisfied either by setting the mass density to zero, which reflects a fluid made of massless carriers (e.g. photons) in a macroscopically non-relativistic regime, or by demanding that  $\theta^{\ast} $ vanishes. This alternative amounts to having an exact clock form (cf. Eq. \eqref{def21eqgal}), $\uptau =-\upmu = \text{d}t_{\text{N}}$, with $t_{\text{N}}$ the absolute Newtonian time, and corresponds to the instance of ordinary fluids.
\end{itemize}

\subsection*{Hydrodynamic-frame transformations}

We would like to now explore the behaviour of the Galilean system when the velocity field undergoes a local Galilean boost.  For kinematics, we use the transformations of $\text{u}$ and $\ast\text{u}$ displayed in Eqs. \eqref{locLor}, combined with the large-$c$ expansions \eqref{uvecexp} for the vectors and \eqref{ufexp} for the forms. It is also wise to keep a subleading term with respect to $c$ in the rapidity 
\begin{equation}
\label{psiexp}
\psi^\ast
=\frac{A}{c}+ \frac{A^{(2)}}{c^3}+\cdots,
\end{equation}
where $A^{(2)}$ is a first-order polynomial in $A$. Trading the index L for G in the variations to stress the shift from Lorentz to Galilei, we find: 
\begin{equation}
\label{deltaGv}
\delta_{\text{G}}\upupsilon= A \ast\! \upupsilon, \quad \delta_{\text{G}} \ast\! \upupsilon=0, \quad \delta_{\text{G}}  \upmu=0 , \quad \delta_{\text{G}} \ast\! \upmu=A \upmu,
\end{equation}
which are indeed the defining properties of Galilean boosts (cf. the already quoted literature). The expansions transform accordingly:
$\delta_{\text{G}}\theta= A \theta^\ast+  \ast \upupsilon(A)$ and 
$\delta_{\text{G}}  \theta^\ast=0$.\footnote{The next-order parameter $A^{(2)}$ enters as follows:  $\delta_{\text{G}}\upupsilon^{(2)}= A \ast\! \upupsilon^{(2)}+A^{(2)}\ast\! \upupsilon$, $ \delta_{\text{G}} \ast\! \upupsilon^{(2)}=A \upupsilon$,  $\delta_{\text{G}}  \upmu^{(2)}=A  \ast\! \upmu$,  $\delta_{\text{G}} \ast\! \upmu^{(2)}=A\upmu^{(2)} + A^{(2)}\upmu$ as well as $\delta_{\text{G}}\theta^{(2)}= A \theta^{\ast(2)}+  \ast \upupsilon^{(2)}(A)+A^{(2)} \theta^{\ast}+  \ast \upupsilon\big(A^{(2)}\big)$ and $\delta_{\text{G}}  \theta^{\ast(2)}= A \theta+  \upupsilon(A)$. \label{deltas2gal}} 
Regarding the observables, Eqs.~\eqref{locLoremt}, \eqref{varlimga} and \eqref{psiexp} yield
\begin{equation}
\label{locgalobs}
\delta_{\text{G}}\varrho =0,
\quad
\delta_{\text{G}} \tilde e= \delta_{\text{G}} (p+\sigma)
=-2 A \tilde\zeta,
\quad
\delta_{\text{G}} \tilde\zeta
=- A \varrho,
\quad
\delta_{\text{G}} \zeta
=-A(\tilde e +p+\sigma)-A^{(2)} \varrho
.
\end{equation}
Finally, the Galilean hydrodynamic system \eqref{galE}, \eqref{galM}, \eqref{galC} and \eqref{galN} transforms as
\begin{equation}
\label{locgalhydeq}
\delta_{\text{G}}\mathcal{E} =-A\mathcal{M}-A^{(2)}\mathcal{N} ,
\quad
\delta_{\text{G}} \mathcal{M}= -A\mathcal{C} 
,
\quad
\delta_{\text{G}} \mathcal{C} 
=-A\mathcal{N} ,
\quad
\delta_{\text{G}} \mathcal{N} =0,
\end{equation}
and is invariant on-shell. 

\subsection*{Physics in the Zermelo frame}

The hydrodynamic invariance under boosts acting on the fluid velocity field is puzzling and contradicts our classical intuition, according to which the velocity field, being a measurable observable, cannot be arbitrarily modified as in relativistic systems. Furthermore, the fluid equations \eqref{galE}, \eqref{galM} and \eqref{galC}, bear little resemblance to the usual energy-balance, Euler and continuity equations. 
 To reach a better understanding of this dynamics it is convenient to work with a concrete parameterization of the sort \eqref{genv}, \eqref{genf},  \eqref{genmetr} and \eqref{gencometr}, well suited to the Galilean limit. This is the Zermelo frame, where the Lorentzian metric and cometric assume the following form:
\begin{equation}
\label{Zmetrcometr}
\text{d}s^2
=-c^2 \Omega^2  \text{d}t^2
+a
\left(\text{d}x-w^x\text{d}t\right)^2,\quad
\partial_s^2
=-\frac{1}{c^2\Omega^2}\left(\partial_t+w^x\partial_x\right)^2
+\frac{1}{a}\partial_x^2.
\end{equation}
In these expressions, all functions depend on $t$ and $x$, and the dependence with respect to the speed of light is explicit. The fluid velocity and the transverse vector are, as in  \eqref{genv},\footnote{Expressions as $\mathbf{b}\cdot \mathbf{c}$ for $\mathbf{b}=b^x\partial_x$ and $\mathbf{c}=c^x\partial_x$ stand for $a_{ij}b^i c^j= a b^x c^x$, not to be confused with $\text{u}\cdot \text{z}=g_{\mu\nu}u^\mu z^\nu$ for $\text{u}=u^\mu\partial_\mu$ and $\text{z}=z^\mu\partial_\mu$. 
\label{convec}}
\begin{equation}
\label{zeru}
\text{u} =\gamma\left(\partial_t+v^x\partial_x\right),
\quad
\ast\text{u} = \frac{\gamma}{c\Omega \sqrt{a}}
\left[\left(c^2\Omega^2 +\mathbf{w}\cdot\left(
\mathbf{v}-\mathbf{w}
\right)
\right)\partial_x
+(v_x-w_x)
\partial_t
\right],
\end{equation}
with the Lorentz factor
\begin{equation}
\gamma=
\frac{1}{\Omega\sqrt{1-\frac{\upalpha^2}{c^2}}}, \quad
\upalpha=\frac{\mathbf{v}-\mathbf{w}}{\Omega}=\alpha^x\partial_x, 
\end{equation}
and similarly for the forms. The rapidity of local Lorentz boosts \eqref{finiteLor} is parameterized as
$\cosh\psi^\ast=
\nicefrac{1}{\sqrt{1-\nicefrac{\mathbf{A}^2}{c^2}}}$ with $\mathbf{A}=A^x(t,x) \partial_x
$,
acting on $\mathbf{v}$ via $\upalpha$ ($\mathbf{w}$ is invariant as is all geometric data) in the celebrated relativistic-composition fashion:
\begin{equation}
\label{alphaprime}
\upalpha' = 
\frac{\upalpha+\mathbf{A}}{1+\frac{\upalpha\cdot\mathbf{A}}{c^2}}
.
\end{equation}

The Galilean limit $c\to \infty$ is straightforward and we reach unambiguously 
\begin{equation}
\label{galuppsilon}
\upupsilon=\frac{1}{\Omega}\hat\partial_t+\alpha^x \partial_x, \quad \ast\upupsilon=\frac{1}{\sqrt{a}}\partial_x, \quad \upmu=-\Omega \text{d}t,\quad \ast\upmu=\sqrt{a}\left(\hat{\text{d}}x-\alpha^x \Omega\text{d}t\right),
\end{equation}
where\footnote{For completeness, we supply 
$\upupsilon^{(2)}=\frac{\upalpha^2}{2}\upupsilon$, $\ast\upupsilon^{(2)}=\frac{\alpha_x}{\sqrt{a}} \upupsilon-\frac{\upalpha^2}{2}\ast\!\upupsilon $,
$\upmu_{(2)}=\frac{\alpha_x}{\sqrt{a}} \ast\!\upmu-\frac{\upalpha^2}{2}\upmu  $ and
$\ast\upmu_{(2)}=\frac{\upalpha^2}{2}\ast\!\upmu $.
From these data, one determines $\theta^{(2)}= \frac{1}{2\Omega \sqrt{a}}\left[\partial_t\left(\sqrt{a}\upalpha^2\right)
+\partial_x\left(\sqrt{a}\upalpha^2v^x\right)\right]=\frac{\upalpha^2}{2}\theta+\frac12\upupsilon\left(\upalpha^2\right)
$ and $\theta^{\ast(2)}=\frac{1}{\Omega \sqrt{a}} \left[\partial_t \alpha_x
+\partial_x\left(\mathbf{w}\cdot\upalpha +\frac{\Omega \upalpha^2}{2}
\right)\right]$, alternatively obtained by the expanding $\Theta$ and $\Theta^\ast$ in powers of $c^{-1}$.\label{theta2gal}} 
\begin{equation}
\hat\partial_t=\partial_t+w^x\partial_x,\quad
\hat{\text{d}}x=\text{d}x -w^x\text{d}t
\end{equation}
with 
\begin{equation}
\label{thetathetaastgal}
\theta=\frac{1}{\Omega}\left(\partial_t
\ln \sqrt{a}+\hat\nabla_x v^x
\right)
,
\quad
\theta^\ast=\frac{1}{\sqrt{a}}\partial_x
\ln \Omega.
\end{equation}
The interpretation is simple. The emerging Newton--Cartan spacetime has a degenerate cometric $ \partial_\ell^2
=\frac{1}{a}\partial_x^2$. It has a natural one-dimensional spatial section with cometric $ \partial_\ell^2
=\frac{1}{a}\partial_x^2$, metric $\text{d}\ell^2=a \hat{\text{d}}x^2$ (which raise and lower the spatial indices) and a Levi-Civita connection $\hat \nabla$. The
fluid  velocity with respect to the coordinate frame inside the spatial section is $\mathbf{v}=v^x \partial_x$. An inertial frame exists moving with velocity $\mathbf{w}=w^x \partial_x$, so that $\upalpha$ is the velocity of the fluid relative to this inertial frame and measured using an invariant time.\footnote{In the Lorentzian spacetime \eqref{Zmetrcometr}, $\gamma \hat\partial_t$  is tangent to a geodesic congruence, when $\Omega$ is a function of $t$ only (see \cite{BigFluid}). This congruence thus defines a locally inertial frame, and keeps this status in the Galilean limit, where it becomes $\upupsilon$. This is why $\upupsilon$ is called the field of observers: it is a necessary tool for carrying out measurements, which comes on top of the defining ingredients for a Newton--Cartan geometry.} The latter is universal if the clock form $\uptau=\Omega \text{d}t$ is exact, which occurs when $\Omega=\Omega(t)$ --- as it is assumed a priori in Refs. \cite{BigFluid, CMPPS1}.

One can work out in the Zermelo frame the large-$c$ expansions of the relativistic equations in coordinate basis. This results in  $c\Omega\nabla_\mu T^{\mu 0}=c^2 \mathcal{C}^{\text{Z}}+\mathcal{E}^{\text{Z}}+\cdots $ and  $\nabla_\mu T^{\mu}_{\hphantom{\mu}x}=c^2 \mathcal{N}^{\text{Z}}_x+\mathcal{M}^{\text{Z}}_x+\cdots $ with
\begin{eqnarray}
\mathcal{E}^{\text{Z}} &=&\left(\frac{1}{\Omega}\hat\partial_t+\theta^w\right)\Pi + \left(\hat\nabla_x+2\varphi_x\right)\Pi^x
+ \theta^w \Pi^x_{\hphantom{x}x},
\label{galEZ}
\\
\label{galMZ}
\mathcal{M}^{\text{Z}}_x &=&\left(\frac{1}{\Omega}\hat{\text{D}}_t+2\theta^w\right)P_x+ \left(\hat\nabla_x+\varphi_x\right)\Pi^x_{\hphantom{x}x}+ \varphi_x \Pi 
,
\\
\label{galCZ}
\mathcal{C}^{\text{Z}} &=&\left(\frac{1}{\Omega}\hat\partial_t+\theta^w\right)\varrho+\left(\hat\nabla_x+2\varphi_x\right)P^x,
\\
\label{galNZ}
\mathcal{N}^{\text{Z}}_x &=&\varphi_x  \varrho.
\end{eqnarray}
Equations resulting from \eqref{galEZ}, \eqref{galMZ} and \eqref{galCZ} are ostensibly heat-transfer, Euler and continuity equations, generalizing on curved spacetime the standard fluid equations displayed e.g. in \cite{Landau}, \S\S 49, 15 and 1. They describe the evolution of the Galilean fluid momenta emerging in the large-$c$ expansion of various components of~$T_{\mu\nu}$: $\Pi$, $\Pi_x$, $P_x$, $\Pi_{xx}$ and $\varrho$. 
\begin{itemize}

\item The total energy density $\Pi= \varrho\left(e+\frac{\upalpha^2}{2}\right)+\mathbf{n}\cdot \upalpha$ features the mass density $\varrho$, the specific energy per unit mass $e$, the \emph{loss/gain current} $\mathbf{n}$ and the inertial velocity $\upalpha$. These appear in the expansions of the relativistic energy density  $\varepsilon= c^2 \varrho +   \varrho \left(e-
\frac{\upalpha^2}{2} \right)-\mathbf{n}\cdot \upalpha+\text{O}\left(\nicefrac{1}{c^2}\right)$ and of the relativistic heat current $q_x= c^2 n_x+ \mathbf{n}\cdot \upalpha\, \alpha_x+Q_x+\text{O}\left(\nicefrac{1}{c^2}\right)$ (see the expression of $T_{\mu\nu}$ displayed in footnote~\ref{Tmunu}). The latter exhibits the classical heat current $Q_x$ responsible for irreversible heat conduction phenomena.

\item The energy flux $\Pi_x=\alpha_x\left[\varrho\left(h +\frac{\upalpha^2}{2}  \right)
+\frac32 \mathbf{n}\cdot \upalpha
\right]+Q_x - \alpha^x \Sigma_{xx}$ accounts also for irreversible phenomena due to the classical stress  $\Sigma_{xx}$, appearing in $\tau_{xx}=-\Sigma_{xx}+\text{O}\left(\nicefrac{1}{c^2}\right)$ ($h=e+\nicefrac{p}{\varrho}$ is the specific enthalpy).

\item The matter current $P_x=\varrho \alpha_x + n_x$ contains an extra contribution with respect to the usual density times velocity term: $n_x$, which we have called loss/gain current. As mentioned previously, this field originates from the leading term in the relativistic heat current $q_x$. It is usually absent and mimics phenomenologically production or destruction phenomena.  

\item The energy--stress tensor $\Pi_{xx}=\varrho \alpha_x \alpha_x+pa-\Sigma_{xx}+2 \alpha_x n_x $ also receives irreversible contributions from the stress and the loss/gain current.

\end{itemize}
On the technical side, the above equations contain the spatial Levi-Civita connection $\hat \nabla $ as well as the time covariant derivative $\hat{\text{D}}_t$, defined for a one-form as $\left(\frac{1}{\Omega}\hat{\text{D}}_t+\theta^w\right)P_x=\frac{1}{\Omega}\left(\hat\partial_t P_x+P_x\partial_x w^x\right)$ with
\begin{equation}
\label{thetaw}
\theta^w=\theta-\left(\hat\nabla_x+\varphi_x\right)\alpha^x, \quad \varphi_x=\sqrt{a}\theta^\ast,
\end{equation}
where $\theta$ and $\theta^\ast$ are given in \eqref{thetathetaastgal}. A wealth of details regarding the Zermelo frame and Galilean connections are available  in Refs. \cite{BigFluid, CMPPS1}, for arbitrary dimension but with $\Omega$ purely time-dependent. 

We should add that the above hydrodynamic system of equations requires a complementary set of constitutive relations expressing $n_x$, $Q_x$ and $\Sigma_{xx}$ in terms derivatives of the velocity field and of local-equilibrium thermodynamic variables. 

Expressions \eqref{galEZ}, \eqref{galMZ}, \eqref{galCZ} and \eqref{galNZ} are related to  \eqref{galE}, \eqref{galM}, \eqref{galC} and \eqref{galN}, reached by starting with the relativistic equations in the generic $\{\text{u}, \ast \text{u} \}$ basis:
\begin{equation}
\label{galtoZ}
\begin{pmatrix}
\mathcal{E} \\
\mathcal{M} \\
\mathcal{C} \\
\mathcal{N}
\end{pmatrix}
=
\begin{pmatrix}
1&-\alpha&\frac{\alpha^2}{2}&-\frac{\alpha^3}{2}\\
0&1&-\alpha&\frac{\alpha^2}{2}&\\
0&0&1&-\alpha\\
0&0&0&1
\end{pmatrix}
\begin{pmatrix}
\mathcal{E}^{\text{Z}} \\
\mathcal{M}^{\text{Z}} \\
\mathcal{C}^{\text{Z}} \\
\mathcal{N}^{\text{Z}}
\end{pmatrix},
\end{equation}
where we have introduced for convenience the following scalars:
\begin{equation}
\label{galZ}
\mathcal{M}^{\text{Z}} =\frac{1}{\sqrt{a}}\mathcal{M}^{\text{Z}}_x, \quad \mathcal{N}^{\text{Z}} =\frac{1}{\sqrt{a}}\mathcal{N}^{\text{Z}}_x,  \quad 
\alpha= \sqrt{a} \alpha^x.
\end{equation}
 The physical observables match likewise: 
\begin{equation}
\label{galrel}
\tilde e= \varrho \left(e-
\frac{\upalpha^2}{2} \right)-\mathbf{n}\cdot \upalpha ,\quad\tilde\zeta=\sqrt{a}n^x,\quad\zeta=\sqrt{a}\left(Q^x+\frac{\upalpha^2}{2}n^x\right), \quad\sigma=-\Sigma^x_{\hphantom{x}x}.
\end{equation}

As advertised earlier, the fluid equations in the Zermelo frame look more familiar and allow to better appreciate the issue of local boosts performed on the fluid velocity field. Considering infinitesimal transformations (small $A$ in \eqref{psiexp}), and computing  $\delta_{\text{G}} \upupsilon^{(2)}$ using $\upupsilon^{(2)}=\frac{\upalpha^2}{2}\upupsilon$ given in footnote \ref{theta2gal} and \eqref{deltaGv}, we find upon comparison with its generic expression displayed in footnote \ref{deltas2gal}
\begin{equation}
A=\sqrt{a} A^x, \quad A^{(2)}= A \upalpha^2.
\end{equation}
We have also used  \eqref{alphaprime}, which provides the familiar expression for the velocity shift in the infinite-$c$ limit
\begin{equation}
\delta_{\text{G}}\alpha^x = A^x.
\end{equation}
Using the relationships \eqref{galrel} between the observables in abstract Cartan and Zermelo frames together with the transformations \eqref{locgalobs}, we reach
\begin{eqnarray}
\label{delgalpaik}
\delta_{\text{G}} p-\delta_{\text{G}} \Sigma^x_{\hphantom{x}x}&=&-2A^x n_x ,
\\
\label{ndelgalQi}
\delta_{\text{G}} Q^x&=&A^x\left(\Sigma^x_{\hphantom{x}x}-\varrho h\right),
\\
\label{ndelgale}
\varrho\delta_{\text{G}} e&=&-A^xn_x,
\\
\label{delgalnk}
\delta_{\text{G}}  n^x&=&-A^x\varrho ,
\end{eqnarray}
on top of $\delta_{\text{G}} \varrho=0$. These leave the Galilean momenta  $\Pi$, $\Pi_x$, $P_x$ and $\Pi_{xx}$
invariant, as they should (recall that the Galilean momenta are the large-$c$ contributions to $T_{\mu\nu} $, which are invariant by definition of a local hydrodynamic-frame transformation).  In turn, Eqs.  \eqref{galEZ}, \eqref{galMZ}, \eqref{galCZ} and \eqref{galNZ} are invariant, in agreement with the transformation laws \eqref{locgalhydeq} and the relations  \eqref{galtoZ}.

From Eq. \eqref{delgalnk}, we learn that the loss/gain current $n_x$ \emph{is not} invariant and this could have been anticipated: for the matter current $P_x=\varrho \alpha_x + n_x$ to be invariant with $\delta_{\text{G}}\alpha^x = A^x$ and $\varrho $ invariant, $n_x$ must undergo an appropriate transformation. However, in standard non-relativistic hydrodynamics $n_x=0$ ($\tilde\zeta=0$ in an abstract Cartan frame), which is a \emph{non-invariant condition} and this ruins the original and genuine relativistic hydrodynamic-frame invariance. This observation has been previously made in \cite{BigFluid}, but we insist on recasting it here as part of the duality relationship operating between one-dimensional Galilean and Carrollian hydrodynamics, to which we will now turn.

\section{Carrollian fluids}\label{carfl}

\subsection*{Carroll structures  and Carrollian  hydrodynamics}

The zero-$c$  limiting relativistic hydrodynamics is the realm of Carrollian fluids. The spacetimes on which they flow are Carrollian manifolds \cite{Duval:2014uva, Duval:2014lpa,Morand:2018tke, Ciambelli:2019lap}. By definition,  Carrollian geometries have a \emph{degenerate metric} $\text{d}\ell^2$ with kernel 
a field of observers  $\upupsilon$. The dual of the field of observers is again the clock form $\uptau$, known also as the Ehresmann connection. Carrollian geometries can be reached at  zero $c$ from Lorentzian spacetimes, as in Eq. \eqref{genmetr}. 

Carrollian hydrodynamics is unraveled through a careful analysis of the Lorentzian equations at vanishing-$c$ limit, or in a complementary fashion, from first principles implemented directly in Carrollian geometries \cite{BigFluid,CMPPS1, Hartong:2015xda, dutch}. We will here adopt the limiting method of  \cite{BigFluid,CMPPS1}, which embraces several orders (doubling the number of equations, as for the Galilean instance), starting with an abstract Cartan frame as in \cite{Campoleoni:2018ltl, CMPR, CMPPS2, Campoleoni:2023fug}. 

Following the Galilean paradigm of Sec. \ref{galfl} and the general expressions \eqref{genv} and \eqref{genf},  we expand the basis
vectors and forms $\{\text{u}, \ast \text{u} \}$ at small $c$:
\begin{eqnarray}
\label{uvecexpc}
&\text{u} = \upupsilon +c^2\upupsilon^{(2)}+\cdots, \quad \ast\text{u} = c\left(\ast\upupsilon +c^2\ast\!\upupsilon^{(2)}+\cdots\right),&
\\
\label{ufexpc}
&\text{u} = c^2 \upmu +c^4\upmu_{(2)}+\cdots,\quad \ast\text{u} = c\left(\ast\upmu +c^2\ast\! \upmu_{(2)}+\cdots\right).&
\end{eqnarray}
The form--vector duality relations remain as in \eqref{dualgal}  and footnote \ref{dual2},
and we find the following behaviour for the metric:
\begin{equation}
\text{d}\ell^2=\lim_{c\to \infty} \text{d}s^2=\ast \upmu^2.
\end{equation}
This bilinear form is degenerate with kernel the field of observers $\upupsilon$, clock-form $\uptau=-\upmu$, and ruler form $\ast \upmu $.
 The spacetime at hand is Carrollian, and $\{\upupsilon, \ast \upupsilon \}$  span the tangent space, while the cotangent space is generated by $\{\upmu, \ast \upmu \}$. Again, the stars are no longer a prescription for Hodge duality, and the extra vectors and forms $\left\{\upupsilon^{(2)}, \ast \upupsilon^{(2)} \right\}$, $\left\{\upmu_{(2)}, \ast \upmu_{(2)} \right\}$ carry further information, inescapable in the spirit of describing the continuous-medium four-equation system.

For the expansions of $\Theta$ and $\Theta^\ast$ we obtain similarly:
\begin{equation}
\Theta=\theta+c^2\theta^{(2)}+\cdots, \quad \Theta^\ast = c\left(\theta^\ast+c^2 \theta^{\ast(2)}+\cdots\right).
\end{equation}
The various functions emerging here appear in the exterior differentials as in \eqref{def21eqgal}, \eqref{def21eqgal2} and \eqref{def21eqgal2ast}.

Similarly to the Galilean case, the small-$c$ expansions of  $\mathcal{L}=0$ and $\mathcal{T}=0$ require a prescription for the behaviour of the physical observables (energy density, heat current and stress tensor). Following \cite{Campoleoni:2018ltl, CMPR, BigFluid, CMPPS1}, we assume that the pressure $p$ and the energy density $\varepsilon$ are of order $1$ in the  limit and keep the same notation. For the heat current and the stress we adopt
\begin{equation}
\chi= \frac{\tilde\zeta}{c}+c \zeta + \text{O}\left(c^3\right),
 \quad \tau=\frac{\sigma}{c^2}+\tilde\sigma + \text{O}\left(c^2\right).
 \label{varlimcar}
\end{equation}
The physical interpretation of the transport functions $\zeta$, $\tilde\zeta$, $\sigma$ and $\tilde\sigma$ is beyond current understanding because a satisfactory theory of Carrollian thermodynamics is unavailable despite some attempts \cite{carrollianlegends}.\footnote{In flat-holography applications, these functions are better understood \cite{Campoleoni:2018ltl, CMPR}. In three dimensions, they provide a boundary materialization of four-dimensional gravitational radiation \cite{CMPPS2, Campoleoni:2023fug}.} Nevertheless, the function $\sigma$ in particular features prominently in the companion papers \cite{APST2,PST3}, which we call the \emph{Carrollian stress}.

Using the above toolkit, $\mathcal{L}$ and $\mathcal{T}$ given in Eqs. \eqref{T-cons-el-mag-nc-force} feature the following small-$c$ expansions:  
\begin{equation}
\mathcal{L}=- \frac{\mathcal{F}}{c^2} -\mathcal{E} + \text{O}\left(c^2\right)
,
\quad
\mathcal{T}= \frac{\mathcal{H}}{c^2}+\mathcal{G} + \text{O}\left(c^2\right)
\label{eqlimcar}
\end{equation}
with 
\begin{eqnarray}
\mathcal{E} &=&-\upupsilon(\varepsilon)- \theta(\varepsilon+p+\tilde\sigma) -\ast\upupsilon\left(\tilde\zeta\right)- 2\theta^\ast \tilde\zeta-\theta^{(2)}  \sigma
,
\label{carE}
\\
\label{carG}
\mathcal{G} &=&\ast\upupsilon(p+\tilde\sigma)+ \theta^\ast (p+\tilde\sigma+\varepsilon)
+ \upupsilon(\zeta)+ 2\theta\zeta
\nonumber\\
&&+ \ast\!\upupsilon^{(2)}(\sigma)+ \theta^{\ast(2)} \sigma
+ \upupsilon^{(2)}\left(\tilde\zeta\right)+ 2\theta^{(2)} \tilde\zeta
,
\\
\label{carF}
\mathcal{F} &=&-\theta \sigma ,
\\
\label{carH}
\mathcal{H} &=&\ast\upupsilon (\sigma)+ \theta^{\ast}  \sigma
+ \upupsilon\left(\tilde\zeta\right)+ 2\theta\tilde\zeta.
\end{eqnarray}
The requirement that $\mathcal{L}$ and $\mathcal{T}$ in \eqref{eqlimcar} vanish for all $c$ and taking the limit $c\to 0$ leads to four equations:
\begin{itemize}
\item The energy equation $\mathcal{E} =0$;
\item The momentum equation $\mathcal{G} =0$;
\item The equation $\mathcal{F} =0$, satisfied either for zero $\sigma$, which is a physical requirement, or in the absence of the Carrollian expansion $\theta$, which is a geometric constraint. The latter  is met for an exact ruler form (see \eqref{def21eqgal}), and is tantamount to having absolute Carrollian space defined in  $\ast\upmu=\text{d}x_{\text{C}}$;

\item An extra equation $\mathcal{H} =0$.

\end{itemize}

\subsection*{Hydrodynamic-frame transformations}

Using the transformation rules \eqref{locLor} in combination with the small-$c$ expansions \eqref{uvecexpc} and \eqref{ufexpc}, and expanding the rapidity as 
\begin{equation}
\label{psiexpc}
\psi^\ast
=c B+ c^3 B^{(2)}+\cdots,
\end{equation}
we find the effect of Carrollian boosts:
\begin{equation}
\label{deltaCv}
\delta_{\text{C}} \upupsilon=0, \quad\delta_{\text{C}}\ast\!\upupsilon= B  \upupsilon, \quad \delta_{\text{C}} \upmu=B\ast\!  \upmu, \quad  \delta_{\text{C}}  \ast\! \upmu=0 ,
\end{equation}
as well as $\delta_{\text{C}}  \theta=0$
and 
$\delta_{\text{C}}\theta^\ast= B \theta+  \upupsilon(B)$.\footnote{The next-order parameter $B^{(2)}$ enters in the higher order corrections to the field of observers and clock form:
 $ \delta_{\text{C}}  \upupsilon^{(2)}=B \ast\!\upupsilon$,  
  $\delta_{\text{C}} \upmu^{(2)}=B\ast\! \upmu^{(2)} + B^{(2)}\ast\! \upmu$,
 $\delta_{\text{C}}\ast\! \upupsilon^{(2)}= B  \upupsilon^{(2)}+B^{(2)}\upupsilon$, 
   $\delta_{\text{C}}  \ast\! \upmu^{(2)}=B \upmu$,  
 $\delta_{\text{C}}  \theta^{(2)}= B \theta^{\ast}+  \ast\upupsilon(B)$
 and
 $\delta_{\text{C}}\theta^{\ast(2)}= B \theta^{(2)}+  \upupsilon^{(2)}(B)+B^{(2)} \theta+  \upupsilon\big(B^{(2)}\big)$. \label{deltas2car}}  The physical transformation parameter $B$ has dimensions of inverse velocity. This is a distinctive trait of the Carroll group, featuring the duality with the Galilei group. 
The observables $\varepsilon$, $p$, $\sigma$, $\tilde\sigma$, $\zeta$ and  $\tilde\zeta$, transform thus as (we use \eqref{locLoremt}, \eqref{varlimcar} and \eqref{psiexpc})
\begin{equation}
\label{loccarlobs}
\delta_{\text{C}} \varepsilon= \delta_{\text{C}} (p+\tilde\sigma)
=-2 B \tilde\zeta,
\quad
\delta_{\text{C}}\sigma =0,
\quad
\delta_{\text{C}} \zeta
=-B(\varepsilon +p+\tilde\sigma)-B^{(2)} \sigma,
\quad 
\delta_{\text{C}}\tilde\zeta
=- B \sigma,
\end{equation}
whereas the Carrollian  equations \eqref{carE}, \eqref{carG}, \eqref{carF} and \eqref{carH} transform as
\begin{equation}
\label{loccarhydeq}
\delta_{\text{C}} \mathcal{E}= B\mathcal{H} 
,
\quad
\delta_{\text{C}}\mathcal{G} =B\mathcal{E}+B^{(2)}\mathcal{F},
\quad
\delta_{\text{C}} \mathcal{F} =0 ,
\quad
\delta_{\text{C}} \mathcal{H} 
=B\mathcal{F},
\end{equation}
and are  on-shell invariant. 

In the absence of any physical intuition about Carrollian fluids,
the Carrollian hydrodynamic-frame invariance revealed here should be taken at face value. It is even a rather desirable feature from a flat-space holographic viewpoint, where Carrollian hydrodynamics in two spacetime dimensions mirrors on null infinity the dynamics of Einstein's equations in the three-dimensional bulk. There, the local two-dimensional Carroll boost invariance reflects residual bulk diffeomorphisms \cite{Campoleoni:2018ltl, CMPR,Campoleoni:2022wmf}. 

\subsection*{The Papapetrou--Randers frame}

Before moving to a comparison of the Galilean and Carrollian fluid equations, we would like to exhibit an alternative, less abstract parameterization inspired by \eqref{genv}, \eqref{genf},  \eqref{genmetr} and \eqref{gencometr} and adapted to the Carrollian limit. This is the Papapetrou--Randers frame, where 
\begin{eqnarray}
\label{PRmetrcometr}
&\text{d}s^2 =- c^2\left(\Omega \text{d}t-b_{x} \text{d}x
\right)^2+a \text{d}x^2,\quad
\partial_s^2
=-\frac{1}{c^2\Omega^2}\partial_t^2
+\frac{1}{a}\left(\partial_x+b_x\partial_t\right)^2,&
\\
\label{PRu}
&\text{u} =\gamma\left(\partial_t+v^x\partial_x\right),
\quad
\ast\text{u} = \frac{c}{\sqrt{a}\sqrt{1-{c^2}{\upbeta^2}}}
\left[\partial_x
+\frac{\beta_x+b_x}{\Omega}
\partial_t
\right],
&
\end{eqnarray}
with  Lorentz factor (the conventions of footnote \ref{convec} hold here)
\begin{equation}
\gamma=
\frac{1+c^2\upbeta\cdot\mathbf{b}}{\Omega\sqrt{1-{c^2}{\upbeta^2}}}, \quad
\mathbf{v}=\frac{c^2\Omega\upbeta}{1+c^2\upbeta\cdot\mathbf{b}}=v^x\partial_x, 
\end{equation}
where  all functions depend on $t$ and $x$, and the dependence with respect to the speed of light is explicit. 

In the Papapetrou--Randers frame, the rapidity of local Lorentz boosts \eqref{finiteLor} is parameterized by an ``inverse velocity'' $\mathbf{B}=B^x(t,x) \partial_x$ as
$\cosh\psi^\ast=
\nicefrac{1}{\sqrt{1-{c^2}\mathbf{B}^2}}
$,
acting on $\mathbf{v}$ via $\upbeta$ ($\mathbf{b}$ is invariant like the other geometric data), which transforms canonically:
\begin{equation}
\label{betaprime}
\upbeta' = 
\frac{\upbeta+\mathbf{B}}{1+c^2\upbeta\cdot\mathbf{B}}
.
\end{equation}

The Carrollian limit in Eqs. \eqref{uvecexpc}, \eqref{ufexpc} gives now
\begin{equation}
\label{caruppsilon}
\upupsilon=\frac{1}{\Omega}\partial_t, \quad \ast\upupsilon=\frac{1}{\sqrt{a}}\left(\hat\partial_x+\frac{\beta_x}{\Omega} \partial_t\right), \quad \upmu=-\Omega \hat{\text{d}}t+\beta_x\text{d}x,\quad \ast\upmu=\sqrt{a}\text{d}x,
\end{equation}
where\footnote{Notice that 
$\upupsilon^{(2)}=\sqrt{a}\beta^x\ast\!\upupsilon-\frac{\upbeta^2}{2}\upupsilon $,
$\ast\upupsilon^{(2)}=\frac{\upbeta^2}{2}\ast\!\upupsilon$, 
$\upmu_{(2)}=\frac{\upbeta^2}{2}\upmu $ and
$\ast\upmu_{(2)}=\sqrt{a}\beta_x\upmu-\frac{\upbeta^2}{2}\ast\!\upmu  $, while
$\theta^{(2)}=\frac{1}{\Omega \sqrt{a}} \left[\partial_x\left(\Omega \sqrt{a} \beta_x\right)+
\partial_t\left(\sqrt{a} \left(\mathbf{b}\cdot\upbeta +\frac{\upbeta^2}{2}\right)
\right)\right]$ and
 $\theta^{\ast(2)}= \frac{1}{2\Omega \sqrt{a}}\left[\partial_x\left(\Omega\upbeta^2\right)
+\partial_t\left(\upbeta^2\left(\beta_x+b_x\right)\right)\right]=\frac{\upbeta^2}{2}\theta^\ast+\frac12\ast\upupsilon\left(\upbeta^2\right)
$, also reached by expanding $\Theta$ and $\Theta^\ast$ in powers of $c$.\label{theta2car}} 
\begin{equation}
\hat\partial_x=\partial_x+\frac{b_x}{\Omega}\partial_t,\quad
\hat{\text{d}}t=\text{d}t -\frac{b_x}{\Omega}\text{d}x
\end{equation}
with 
\begin{equation}
\label{thetathetaastcar}
\theta=\frac{1}{\Omega}\partial_t
\ln \sqrt{a}
,
\quad
\theta^\ast=\frac{1}{\sqrt{a}}\left(\partial_x
\ln \Omega+\frac{1}{\Omega}\partial_t\left(b_x+\beta_x\right)\right).
\end{equation}
The limiting Carrollian geometry has a degenerate metric $ \text{d}\ell^2
=a\text{d}x^2$. It has a natural one-dimensional spatial section endowed with the metric $\text{d}\ell^2=a \text{d}x^2$
and
cometric $ \partial_\ell^2
=\frac{1}{a}\hat{\partial}_x^2$,  which lower  and raise the spatial indices, and a Levi-Civita connection $\hat \nabla$ with Christoffel symbol $\hat \gamma^x_{xx}=\hat\partial_x\ln a$.

The relativistic equations in the Papapetrou--Randers coordinate basis behave at small $c$ as follows:  $\frac{c}{\Omega}\nabla_\mu T^\mu_{\hphantom{\mu} 0}=\frac{1}{c^2} \mathcal{F}_{\text{PR}}+\mathcal{E}_{\text{PR}}+\ldots $ and  $\nabla_\mu T^{\mu x}=\frac{1}{c^2} \mathcal{H}_{\text{PR}}^x+\mathcal{G}_{\text{PR}}^x+\ldots \, $, with
\begin{eqnarray}
\mathcal{E}_{\text{PR}}&=&-\left(\frac{1}{\Omega}\partial_t+\theta\right)\Pi - \left(\hat\nabla_x+2\varphi_x\right)\Pi^x
- \theta \Pi^x_{\hphantom{x}x},
\label{carEPR}
\\
\label{carGPR}
\mathcal{G}_{\text{PR}}^x&=&\left(\frac{1}{\Omega}\hat{\text{D}}_t+2\theta\right)P^x+ \left(\hat\nabla_x+\varphi_x\right)\Pi^{xx}+ \varphi^x \Pi 
,
\\
\label{carFPR}
\mathcal{F}_{\text{PR}}  &=&-\theta  \tilde \Pi^x_{\hphantom{x}x},
\\
\label{carHPR}
\mathcal{H}_{\text{PR}}^x&=&\left(\frac{1}{\Omega}\hat{\text{D}}_t+2\theta\right)\Pi^x+ \left(\hat\nabla_x+\varphi_x\right)\tilde\Pi^{xx}.
\end{eqnarray}
Equations $\mathcal{E}_{\text{PR}}=0$ and $\mathcal{G}_{\text{PR}}^x=0$ are the energy-balance and the momentum-conservation equations. The extra equation $\mathcal{H}_{\text{PR}}^x=0$ is a sort of continuity equation, though for a higher-spin field rather than a scalar. Together with $\mathcal{F}_{\text{PR}}=0$, they describe the evolution of the  Carrollian momenta appearing in the small-$c$ expansion of $T_{\mu\nu}$: $\Pi$, $\Pi_x$, $P_x$, $\Pi_{xx}$ and $\tilde\Pi_{xx}$. 
Contrary to the Galilean instance, our intuition about these is poor and we can only state their expressions:
\begin{equation}
\label{carmom}
\begin{cases}
\Pi=\varepsilon +2\beta_x Q^x-\beta_x\beta_x\Sigma^{xx}\\
\Pi^x=Q^x-\beta^x\Sigma^{x}_{\hphantom{x}x}\\
P^x=\pi^x-\beta^x\left(\Xi^{x}_{\hphantom{x}x}-p-Q^x\beta_x-\varepsilon\right)+\frac{\upbeta^2}{2}Q^x\\
\Pi^{x}_{\hphantom{x}x}=p-\Xi^{x}_{\hphantom{x}x}+2Q^x\beta_x\\
\tilde\Pi^{x}_{\hphantom{x}x}=-\Sigma^{x}_{\hphantom{x}x}.
\end{cases}
\end{equation}
In these expressions, $Q^x$, $\pi^x$, $\Sigma^{xx}$ and $\Xi^{xx}$ emerge in the small-$c$ expansions of the relativistic heat current and stress tensor:
$q^x= Q^x+ c^2 \pi^x+\text{O}\left(c^4\right)$ and $\tau^{xx}=-\frac{\Sigma^{xx} }{c^2}-\Xi^{xx} + \text{O}\left(c^2\right)$. Following the Lorentzian and Galilean paradigms, these observables are ultimately expressed in the form of  constitutive relations in terms of kinematic and local-equilibrium thermodynamic variables --- bearing in mind the absence of clear definitions of the latter in Carrollian physics.  

As mentioned earlier, $\hat \nabla $ is the Levi-Civita connection on the spatial base of the Carrollian bundle, whereas  $\hat{\text{D}}_t$ stands for a Carrollian time covariant derivative (see \cite{BigFluid, CMPPS1} for more information) acting on a vector as $\frac{1}{\Omega}\hat{\text{D}}_tP^x=\left(\frac{1}{\Omega}\partial_t +\theta\right)P^x$ with $\theta$ given in \eqref{thetathetaastcar}. We have also introduced 
\begin{equation}
\label{phicar}
\varphi_x=\frac{1}{\Omega}\left(\partial_t b_x+\partial_x \Omega\right)=\sqrt{a}\theta^\ast-\frac{1}{\Omega} \partial_t \beta_x,
\end{equation}
where $\theta^\ast$ is displayed in \eqref{thetathetaastcar}. Using these tools, it is straightforward to relate the above 
equations \eqref{carEPR}, \eqref{carGPR}, \eqref{carFPR} and \eqref{carHPR} to those obtained previously from a different Lorentzian basis, \eqref{carE}, \eqref{carG}, \eqref{carF} and \eqref{carH}:
\begin{equation}
\label{cartoPR}
\begin{pmatrix}
\mathcal{G} \\
\mathcal{E} \\
\mathcal{H} \\
\mathcal{F}
\end{pmatrix}
=
\begin{pmatrix}
1&\beta&\frac{\beta^2}{2}&\frac{\beta^3}{2}\\
0&1&\beta&\frac{\beta^2}{2}&\\
0&0&1&\beta\\
0&0&0&1
\end{pmatrix}
\begin{pmatrix}
\mathcal{G}_{\text{PR}} \\
\mathcal{E}_{\text{PR}} \\
\mathcal{H}_{\text{PR}} \\
\mathcal{F}_{\text{PR}}
\end{pmatrix}
\end{equation}
with
\begin{equation}
\label{carPR}
\mathcal{G}_{\text{PR}} =\sqrt{a}\mathcal{G}_{\text{PR}}^x, \quad \mathcal{H}_{\text{PR}} =\sqrt{a}\mathcal{H}_{\text{PR}}^x,  \quad 
\beta= \frac{1}{\sqrt{a}}\beta_x.
\end{equation}
The sets of observables match as follows: 
\begin{equation}
\label{carrel}
\tilde\sigma= -\Xi^x_{\hphantom{x}x}+ \upbeta^2 \Sigma^x_{\hphantom{x}x},\quad\sigma=-\Sigma^x_{\hphantom{x}x}
, \quad \zeta=\sqrt{a}\left(\pi^x-\frac{\upbeta^2}{2}Q^x\right), \quad
\tilde\zeta=\sqrt{a}Q^x.
\end{equation}

We close this chapter with the analysis of local boosts in the Papapetrou--Randers frame, and consider infinitesimal transformations (small $B$ in \eqref{psiexpc}). In the Carrollian limit, the transformation rule \eqref{betaprime} translates into a shift of the inverse velocity:
\begin{equation}
\delta_{\text{C}}\beta^x = B^x.
\end{equation}
Computing  $\delta_{\text{C}} \ast\!\upupsilon^{(2)}$ using $\ast\upupsilon^{(2)}=\frac{\upbeta^2}{2}\ast\!\upupsilon$ given in footnote \ref{theta2car} and \eqref{deltaCv}, we find upon comparison with its generic expression displayed in footnote \ref{deltas2car} the following entries for \eqref{psiexpc}:
\begin{equation}
B=\sqrt{a} B^x, \quad B^{(2)}= B \upbeta^2.
\end{equation}

The transformations \eqref{loccarlobs} of the observables in abstract Cartan frame, combined with the above dictionary \eqref{carrel} lead to the following in Papapetrou--Randers:
\begin{eqnarray}
\label{ndelcare}
\delta_{\text{C}} \varepsilon&=&-2B_xQ^x,
\\
\label{delcarpaik}
\delta_{\text{C}} \Xi^x_{\hphantom{x}x}-\delta_{\text{C}} p&=&2 B^x\left(Q_x+\beta_x \Sigma^x_{\hphantom{x}x} \right),
\\
\label{delcarSig}
\delta_{\text{C}} \Sigma^x_{\hphantom{x}x}&=&0,
\\
\label{delcarQ}
\delta_{\text{C}}  Q^x&=& B^x\Sigma^x_{\hphantom{x}x},
\\
\label{ndelcarpi}
\delta_{\text{C}} \pi^x&=&B^x\left(\Xi^x_{\hphantom{x}x}-p-\varepsilon+ \frac{\upbeta^2}{2}\Sigma^x_{\hphantom{x}x}+\beta_xQ^x \right)
.
\end{eqnarray}
These leave the Carrollian momenta  $\Pi$, $\Pi_x$, $P_x$, $\Pi_{xx}$ and  $\tilde\Pi_{xx}$ invariant, as it happens in the Galilean case. Hence, equations  \eqref{carEPR}, \eqref{carGPR}, \eqref{carFPR} and \eqref{carHPR} are invariant, in line with the transformation laws \eqref{loccarhydeq} and the relations  \eqref{cartoPR}.
Contrary to Galilean fluid dynamics, no quest for breaking the hydrodynamic-frame invariance is needed here on Carrollian physical grounds.

\section{Duality and outlook}\label{dualout}

The stage is now set for comparing Galilean and Carrollian hydrodynamics, and unravelling   their duality relationship. To this end, we would like to summarize the results of Sects. \ref{galfl} and \ref{carfl}.

Galilean, i.e. ordinary non-relativistic fluids, in one spatial dimension are naturally defined on a two-dimensional  Newton--Cartan spacetime. Working with an abstract Cartan frame, the basic ingredients are the degenerate cometric $\partial_\ell^2=\ast \upupsilon^2$ and the clock form $\uptau=-\upmu$, as well as the field of observers $\upupsilon$. The dual to the spatial vector $\ast \upupsilon$ is the ruler form $\ast \upmu$, which defines the spatial metric $\text{d}\ell^2=\ast \upmu^2$. The forms $\{\upmu, \ast \upmu \}$ span the cotangent space, and their exterior differentials \eqref{def21eqgal} define $\theta$ and $\theta^\ast$, which carry information on the connection that ultimately equips the geometry.

As a novelty, we have introduced a piece of extra structure on this geometry, supported by a pair of forms  $\left\{\upmu_{(2)}, \ast \upmu_{(2)} \right\}$ and a pair of vectors  $\left\{\upupsilon^{(2)}, \ast \upupsilon^{(2)} \right\}$ obeying the relationships quoted in footnote \ref{dual2}, and providing two further functions, $\theta^{(2)}$ and $\theta^{\ast(2)}$, through their differentials \eqref{def21eqgal2} and \eqref{def21eqgal2ast}. This structure is necessary in order to reach the full set of four equations $\mathcal{E}=0$,  $\mathcal{M}=0$, as well as $\mathcal{C}=0$ and  $\mathcal{N}=0$ (see \eqref{galE}, \eqref{galM}, \eqref{galC} and \eqref{galN}).  The latter involve the physical observables $\tilde e$, $p$ and $\varrho$, together with the transport-like  variables $\zeta$, $\tilde\zeta$ and $\sigma$. 

Alternatively, a Zermelo frame can be used, where the fluid kinematics is captured by a velocity field $\alpha^x$ with respect to a locally inertial frame. The observables are here more physical, being the total energy density $\Pi$, which captures the specific energy $e$, the energy flux $\Pi_x$ including the heat current $Q_x$, the energy--stress tensor $\Pi_{xx}$ including the irreversible stress $\Sigma_{xx}$, and the matter current $P_x$, which also contains the exotic loss/gain current $n_x$. 

The complete system (in Zermelo or Cartan representations) is invariant under local hydrodynamic-frame transformations (Galilean boosts acting on the fluid velocity), as long as the loss/gain current does not vanish. In the natural instance where the latter is zero, this gauge invariance is broken and the fluid velocity becomes physical, i.e. measurable. 

Turning to Carrollian hydrodynamics, the steps are very similar, except that the relevant geometries are now Carrollian. In the general frame, the degenerate metric is $\text{d}\ell^2=\ast \upmu^2$ with field of observers $\upupsilon$ and clock form $\uptau=-\upmu$. The dual to the ruler form $\ast \upmu$ is the spatial vector $\ast \upupsilon$, which provides a spatial cometric  $\partial_\ell^2=\ast \upupsilon^2$. The basis forms define again $\theta$ and $\theta^\ast$, through  \eqref{def21eqgal}. The new geometric features are two functions $\theta^{(2)}$ and $\theta^{\ast(2)}$ derived using  \eqref{def21eqgal2} and \eqref{def21eqgal2ast}. As in the Galilean case, these provide extra information on the Carrollian manifold, and make it possible to reveal the four equations  $\mathcal{E}=0$,  $\mathcal{G}=0$, as well as $\mathcal{F}=0$ and  $\mathcal{H}=0$ --- see \eqref{carE}, \eqref{carG}, \eqref{carF} and \eqref{carH}.  These equations involve the physical observables $\varepsilon$ and $p$, together with the variables $\zeta$, $\tilde\zeta$ and $\sigma$, $\tilde\sigma$. 

In the Papapetrou--Randers frame, an inverse velocity field $\beta^x$ parameterizes the fluid kinematics. Fluid momenta appear as in non-relativistic fluids (the total energy density $\Pi$, the energy flux $\Pi_x$, the energy--stress tensor $\Pi_{xx}$ and the matter current $P_x$), which now
contain two kinds of heat currents, $\pi_x$ and $Q_x$, as well as two sorts of stress tensors $\Xi_{xx}$ and $\Sigma_{xx}$.

Carrollian fluid equations are invariant under local hydrodynamic-frame transformations (Carrollian boosts acting on the fluid velocity), and no privileged situation is obvious that would lead to the breaking of this invariance.
\begin{table}[!ht]
\begin{center}
  \begin{tabular}{|c|c||c|c||c|c|}
    \hline
  Gal  & Car& Gal  & Car  & Gal & Car      \\
    \hline
    \hline
    $\upupsilon$ & $\ast\upupsilon$ &  $\upupsilon^{(2)}$ & $\ast\upupsilon^{(2)}$ & $\theta$ &  $\theta^{\ast}$      \\
    \hline
    $\ast\upupsilon$  & $\upupsilon$ & $\ast\upupsilon^{(2)}$  & $\upupsilon^{(2)}$ & $\theta^{\ast}$ &  $\theta$      \\
\hline 
$-\upmu$& $\ast\upmu$ & $-\upmu_{(2)}$& $\ast\upmu_{(2)}$ & $\theta^{(2)}$ &  $\theta^{\ast(2)}$      \\
    \hline
    $\ast\upmu$& $-\upmu$& $\ast\upmu_{(2)}$& $-\upmu_{(2)}$ & $\theta^{\ast(2)}$ &  $\theta^{(2)}$     \\
\hline
   \end{tabular}
\end{center}
    \caption{Duality among kinematical variables.}\label{cartkin}
\end{table}

The above geometric derivations, featuring in particular new pairs of vectors and forms, and granting the doubling of equations so as to capture e.g. the continuity equation in the Galilean case, is worth generalizing to higher dimensions. The distinctiveness of  two spacetime dimensions is that
Galilean and Carrollian groups are isomorphic \cite{Duval:2014uoa}. They are generated by a unique boost  as well as time and space translations.  The swapping of time and space defines the relationship between Galilei and Carroll groups.  This is also the principle behind the duality relating Galilean and Carrollian hydrodynamics. Indeed,
in a Cartan frame, the kinematical variables are related by the interchange of the starred ones with their non-starred relatives, as summarized in Tab. \ref{cartkin}. 
Under this transformation, the hydrodynamic expressions  \eqref{galE}, \eqref{galM}, \eqref{galC} and \eqref{galN} are mapped onto  \eqref{carE}, \eqref{carG}, \eqref{carF} and \eqref{carH} according to Tab. \ref{carteq}, provided the dynamical variables are transformed following Tab. \ref{cartdyn}.
\begin{table}[!ht]
\begin{center}
  \begin{tabular}{|c|c|}
    \hline
  Gal  & Car      \\
    \hline
    \hline
    $\mathcal{E}$ & $\mathcal{G}$      \\
    \hline
    $\mathcal{M}$  & $-\mathcal{E}$ 
          \\
\hline 
$\mathcal{C}$& $\mathcal{H}$       \\
    \hline
    $\mathcal{N}$& $-\mathcal{F}$    \\
\hline
   \end{tabular}
\end{center}
    \caption{Duality among hydrodynamic expressions.}\label{carteq}
\end{table}
\begin{table}[!ht]
\begin{center}
  \begin{tabular}{|c|c|}
    \hline
  Gal  & Car      \\
    \hline
    \hline
    $\tilde e$ & $p+\tilde\sigma$      \\
    \hline
    $p+\sigma$  & $\varepsilon$ 
          \\
\hline 
$\varrho$& $\sigma$       \\
    \hline
    $\tilde\zeta$& $\tilde\zeta$    \\
\hline $\zeta$& $\zeta$    \\
\hline
   \end{tabular}
\end{center}
    \caption{Duality among dynamical variables.}\label{cartdyn}
\end{table}
This is the essence of the advertised duality for one-dimensional Galilean and Carrollian fluids. This duality also maps the hydrodynamic-frame transformations under local Galilean boosts \eqref{deltaGv}, \eqref{locgalobs} and \eqref{locgalhydeq} onto the Carroll-boost transformations  \eqref{deltaCv}, \eqref{loccarlobs} and \eqref{loccarhydeq}, by interchanging $\left\{A, A^{(2)}\right\}$ with $\left\{B, B^{(2)}\right\}$.

Although it is rooted in the well-established Galilean--Carrollian algebra isomorphism, the above description constitutes an original and concrete realization thereof in the area of fluids, involving complete sets of hydrodynamic equations. Swapping time and space amounts to exchanging longitudinal and transverse directions with respect to the always time-like velocity field. Physically, this means that equilibrium-type observables like mass, energy  or pressure are mapped to or mixed with out-of-equilibrium variables like the irreversible stress or the heat current. In order to better illustrate this statement, 
it is instructive to recast the Galilei--Carroll duality in Zermelo and Papapetrou--Randers language, because this conveys a certain intuition --- at least as far as the Galilean side is concerned. 

The kinematical relationships of Tab. \ref{cartkin} are translated with the help of Eqs. \eqref{galuppsilon}, \eqref{thetathetaastgal}, \eqref{thetaw}, 
\eqref{caruppsilon}, \eqref{thetathetaastcar} and \eqref{phicar}.  The output is summarized in Tab.  \ref{ZPRkin}. 
\begin{table}[!ht]
\begin{center}
  \begin{tabular}{|c|c||c|c||}
    \hline
  Gal  & Car& Gal  & Car     \\
    \hline
    \hline
    $(t,x)$ & $(x,t)$ &  $\hat\partial_t$ & $\hat\partial_x$      \\
    \hline
    $\Omega$  & $\sqrt{a}$ & $\partial_x$  & $\partial_t$       \\
\hline 
$\sqrt{a}$& $\Omega$ & $\theta^w$ & $\sqrt{a}\varphi^x$     \\
    \hline
    $w^x$& $\frac{b_x}{\Omega}$& $\sqrt{a}\alpha^x$& $\sqrt{a}\beta^x$    \\
\hline
   \end{tabular}
\end{center}
    \caption{Kinematical duality in Zermelo/Papapetrou--Randers.}\label{ZPRkin}
\end{table}
Likewise, the content of Tab. \ref{cartdyn}, combined with Eqs. \eqref{galrel} and 
\eqref{carrel} delivers  Tab. \ref{ZPRdyn},  translating at the same time a simple pairing of Galilean with Carrollian momenta, all invariant under their respective boosts. The duality relationships among the Galilean hydrodynamic expressions \eqref{galEZ},
\eqref{galMZ},
\eqref{galCZ},
\eqref{galNZ}, 
and their Carrollian relatives 
\eqref{carEPR},
\eqref{carGPR},
\eqref{carFPR},
\eqref{carHPR},
are obtained using \eqref{galtoZ} and \eqref{cartoPR} in Tab. \ref{carteq}. They are reported in Tab. \ref{ZPReq}.
\begin{table}[!ht]
\begin{center}
  \begin{tabular}{|c|c||c|c|}
    \hline
  Gal  & Car   & Gal & Car   \\
    \hline
    \hline
    $\sqrt{a}n^x$ & $\sqrt{a}Q^x$ &  $\Pi$ & $\Pi^{x}_{\hphantom{x}x} $ \\
    \hline
    $\sqrt{a}\left(Q^x+\frac{\upalpha^2}{2}n^x\right)$  & $\sqrt{a}\left(\pi^x-\frac{\upbeta^2}{2}Q^x\right)$ &$\Pi^{x}_{\hphantom{x}x} $ &   $\Pi$ 
          \\
\hline 
$\varrho \left(e+
\frac{\upalpha^2}{2} \right)-\mathbf{n}\cdot \upalpha $& $p-\Xi^{x}_{\hphantom{x}x}$  &$\sqrt{a}P^x$ & $\sqrt{a}\Pi^x$    \\
    \hline
    $p-\Sigma^{x}_{\hphantom{x}x}$& $\varepsilon$ & $\sqrt{a}\Pi^x$ & $\sqrt{a}P^x$    \\
\hline $\varrho$& $-\Sigma^{x}_{\hphantom{x}x}$    &$\varrho$ &$\tilde\Pi^{x}_{\hphantom{x}x} $\\
\hline
   \end{tabular}
\end{center}
    \caption{Dynamical duals in Zermelo/Papapetrou--Randers.}\label{ZPRdyn}
\end{table}
\begin{table}[!ht]
\begin{center}
  \begin{tabular}{|c|c|}
    \hline
  Gal  & Car      \\
    \hline
    \hline
    $\mathcal{E}^{\text{Z}}$ & $\mathcal{G}_{\text{PR}}$      \\
    \hline
    $\mathcal{M}^{\text{Z}}$  & $-\mathcal{E}_{\text{PR}}$ 
          \\
\hline 
$\mathcal{C}^{\text{Z}}$& $\mathcal{H}_{\text{PR}}$       \\
    \hline
    $\mathcal{N}^{\text{Z}}$& $-\mathcal{F}_{\text{PR}}$    \\
\hline
   \end{tabular}
\end{center}
    \caption{Duality in fluid equations  in Zermelo/Papapetrou--Randers.}\label{ZPReq}
\end{table}

Several remarks are in order here. First, we see in Tab. \ref{ZPRdyn} that the Galilean loss/gain current $n^x$ is mapped to a Carrollian heat current $Q^x$. In ordinary Galilean fluids $n^x=0$ and this breaks the local Galilean hydrodynamic-frame invariance, and confers a physical meaning on the velocity $\alpha^x$. A similar role is manifestly played in Carrollian fluids by $Q^x$.  

We learn from Tab. \ref{ZPRdyn} again that the Galilean mass density $\varrho$ is dual to the Carrollian stress $\tilde\Pi^{x}_{\hphantom{x}x} =-\Sigma^{x}_{\hphantom{x}x}$. Similarly, the Carrollian energy density $\varepsilon$ is mapped onto  the total Galilean stress $p-\Sigma^{x}_{\hphantom{x}x}$. This echoes the more general comment made earlier regarding the sought after duality relationship, expected to be of the equilibrium/out-of-equilibrium type, reminiscent of a strong/weak-coupling type of duality. Even though the duality at hand is circumscribed to two spacetime dimensions, it conveys features that seem  emblematic of Carrollian physics. These features may explain why it is difficult to make sense of thermodynamics in Carrollian systems, when the working perspective is deeply Galilean as in Ref.  \cite{carrollianlegends}. Particles might have to be traded for extended objects (as quoted without much progress in \cite{CMPPS1}) and the concept of equilibrium revisited.  

Observe finally that the permutation of time and space makes the time evolution in Carrollian fluids utterly different from that in their Galilean counterparts, mapping a Cauchy problem to a boundary value problem, and vice-versa. We investigate this side of the Carrollian programme from a mathematically rigorous perspective in the companion works \cite{APST2, PST3}, in the framework of flat Newton--Cartan and Carrollian two-dimensional geometries (i.e. with $\Omega=a=1$ and $w^x=b_x=0$ in  Zermelo and Papapetrou--Randers frames). We conclude by noting that in this case the quantities $\theta$ and $\varphi_x$ vanish, the equation $\mathcal{F}_{\text{PR}} = 0$ trivializes, and, assuming $Q^x=0$, the remaining equations \eqref{carEPR}, \eqref{carGPR} and \eqref{carHPR} take the form
\begin{align}
    \label{flat_carEPR}
    \mathcal{E}_{\text{PR}} &=- \partial_t \left(\varepsilon + \beta^2 \sigma\right) - \partial_x (\beta \sigma) = 0, \\ 
    \label{flat_carGPR}
    \mathcal{G}_{\text{PR}} & = \partial_t(\beta(\varpi+\varepsilon)  + \pi ) + \partial_x \varpi = 0,\\
    \label{flat_carHPR}
    \mathcal{H}_{\text{PR}} & = \partial_t (\beta \sigma) + \partial_x \sigma = 0,
\end{align}
where we have used \eqref{carPR}, \eqref{carrel} and the notation
\[ \varpi = p - \Xi^x{}_x, \qquad \pi = \pi^x. \]
In fact, in \cite{APST2,PST3} we introduce a notion of \emph{isentropic} Carrollian equations to moreover trivialize equation \eqref{flat_carGPR} in the case $\pi = 0$, thus working in the case of the vanishing of the entire relativistic heat current $q^x$, at least to order $c^2$. The qualifier ``isentropic'' here refers to a novel concept of \emph{Carrollian entropy} $\tilde s(\varpi, \sigma)$ imitating (or, better, dual to) Galilean entropy $s(e,\varrho)$. The usual Galilean thermodynamic relation $\text{d}(e \varrho)=\varrho T\text{d}s  + h \text{d}\varrho$ would then lead to a Carrollian law
\begin{equation}
\label{carthermo}
\text{d}\varpi=\sigma \tilde T\text{d}\tilde s  + \tilde h \text{d}\sigma,
\end{equation}
defining a Carrollian temperature $\tilde T$ and enthalpy $\tilde h$. All of this is rather formal and undoubtedly calls for further careful examination.

\section*{Acknowledgements}
The ideas elaborated in the present work were triggered during discussions at the \emph{Journ\'ees Relativistes de Tours}, organised by X. Bekaert, Y. Herfray, S. Solodukhin and M. Volkov,  and held at the Institut Denis Poisson in June 2023. Further fruitful exchanges followed during the summer 2024 meeting \emph{Carroll Think Tank: Mathematics and physics}, held at the Aristotle University of Thessaloniki and financed by a \emph{Partenariat Hubert Curien France-Gr\`ece}. Marios Petropoulos and Simon Schulz thank the Department of Pure Mathematics and Mathematical Statistics for its kind hospitality, and the University of Cambridge for financial support. Nikolaos Athanasiou wishes to gratefully acknowledge an H.F.R.I. grant for postdoctoral researchers (3rd call, no. 7126). We are grateful to our colleagues M. Beauvillain, A. Campoleoni, A.~Delfante, B. Oblak, S. Pekar, D. Rivera-Betancour, M. Vilatte for priceless  scientific exchanges.


\begin{thebibliography}{99.}%

\bibitem{Levy}
J.-M. L\'evy-Leblond, \textit{Une nouvelle limite non-relativiste du groupe de Poincar\'e},  
\href{https://eudml.org/doc/75509}{A. Inst. Henri Poincar\'e \textbf{III} (1965) 1.} 

\bibitem{SenGupta}
N. D. Sen Gupta, \textit{On an analogue of the Galilei group},  
\href{https://link.springer.com/content/pdf/10.1007/BF02740871.pdf}{Il Nuovo Cimento \textbf{XLIVA} (1966) 512.} 

\bibitem{Henneaux:1979vn}
M. Henneaux, \textit{Geometry of zero-signature space--times,}
Bull. Soc. Math. Belg. \textbf{31} (1979) 47.

 \bibitem{NC} E. Cartan, \textit{Sur les vari\'et\'es \`a connexion affine, et la th\'eorie de la relativit\'e g\'en\'eralis\'ee} (premi\`ere partie),  \href{http://archive.numdam.org/article/ASENS_1924_3_41__1_0.pdf}{Ann. \'Ecole norm. \textbf{41} (1924) 1}.

\bibitem{Duv}
  C. Duval and P.~A. Horvathy,
  \textit{Non-relativistic conformal symmetries and Newton--Cartan structures,}
  \href{https://iopscience.iop.org/article/10.1088/1751-8113/42/46/465206}{J. Phys. \textbf{A42} (2009) 465206},  \href{https://arxiv.org/abs/0904.0531}{arXiv:0904.0531 [math-ph].}

\bibitem{Duval:2014uoa}
  C. Duval, G.~W. Gibbons, P.~A. Horvathy and P.~M. Zhang,
  \textit{Carroll versus Newton and Galilei: two dual non-Einsteinian concepts of time,}
  \href{https://iopscience.iop.org/article/10.1088/0264-9381/31/8/085016}{Class.\ Quant.\ Grav.\  {\bf 31} (2014) 085016}, \href{http://arxiv.org/abs/1402.0657}{arXiv:1402.0657 [gr-qc].}
  
\bibitem{Duval:2014uva}
  C. Duval, G.~W. Gibbons and P.~A. Horvathy,
  \textit{Conformal Carroll groups and BMS symmetry,}
  \href{https://iopscience.iop.org/article/10.1088/0264-9381/31/9/092001}{Class.\ Quant.\ Grav.\  {\bf 31} (2014) 092001}, \href{http://arxiv.org/abs/1402.5894}{arXiv:1402.5894 [gr-qc].}  

\bibitem{Duval:2014lpa}
  C.~Duval, G.~W.~Gibbons and P.~A.~Horvathy,
 \textit{Conformal Carroll groups,} 
  \href{https://iopscience.iop.org/article/10.1088/1751-8113/47/33/335204}{J.\ Phys. {\bf A47} (2014) 335204},
 \href{http://arxiv.org/abs/1403.4213}{arXiv:1403.4213 [hep-th].} 
 
\bibitem{Bekaert:2014bwa}
  X.~Bekaert and K.~Morand,
  {\it Connections and dynamical trajectories in generalised Newton--Cartan gravity I. An intrinsic view},
  \href{https://aip.scitation.org/doi/10.1063/1.4937445}{J.\ Math.\ Phys.\  {\bf 57} (2016)  022507}, 
  \href{https://arxiv.org/abs/1412.8212}{arXiv:1412.8212 [hep-th].}

  \bibitem{Bekaert:2015xua}
  X.~Bekaert and K.~Morand,
  {\it Connections and dynamical trajectories in generalised Newton--Cartan gravity II. An ambient perspective},
  \href{https://aip.scitation.org/doi/10.1063/1.5030328}{J. Math. Phys. \textbf{59}, no.7, 072503 (2018)}
 \href{https://arxiv.org/abs/1505.03739}{arXiv:1505.03739 [hep-th].}
 
  \bibitem{Morand:2018tke}
  K.~Morand,
  \textit{Embedding Galilean and Carrollian geometries I. Gravitational waves},
  \href{https://aip.scitation.org/doi/10.1063/1.5130907}{J. Math. Phys. \textbf{61}, no.8, 082502 (2020)},
  \href{http://arxiv.org/abs/1811.12681}{arXiv:1811.12681 [hep-th].} 

  \bibitem{Ciambelli:2019lap}
  L.~Ciambelli, R.~G.~Leigh, C.~Marteau and P.~M.~Petropoulos,
  \textit{Carroll structures, null geometry and conformal isometries},
  \href{https://journals.aps.org/prd/abstract/10.1103/PhysRevD.100.046010}{Phys.\ Rev.\  {\bf D100} (2019)  046010},
  \href{https://arxiv.org/abs/1905.02221}{arXiv:1905.02221 [hep-th].}
  
  \bibitem{Herfray:2021qmp}
Y.~Herfray,  \textit{Carrollian manifolds and null infinity: a view from Cartan geometry},
\href{https://iopscience.iop.org/article/10.1088/1361-6382/ac635f}{Class.\ Quant.\ Grav.\  {\bf 39} (2022) 215005},
  \href{https://arxiv.org/abs/2112.09048}{arXiv:2112.09048 [gr-qc]}.


  \bibitem{Bagchi:2010zz}
  A.~Bagchi, \textit{Correspondence between asymptotically flat spacetimes and nonrelativistic conformal field theories},
  \href{https://journals.aps.org/prl/abstract/10.1103/PhysRevLett.105.171601}{Phys. Rev. Lett. \textbf{105} (2010), 171601},   \href{https://arxiv.org/abs/1006.3354}{arXiv:1006.3354 [hep-th].}


\bibitem{APST2}
N. Athanasiou, P. M. Petropoulos, S. M. Schulz and G. Taujanskas, \textit{One-dimensional Carrollian fluids II: $C^1$ blow-up criteria}, CPHT-RR027.052024,
 \href{https://arxiv.org/abs/2407.05971}{arXiv:2407.05971 [math.AP].}

\bibitem{PST3}
P. M. Petropoulos, S. M. Schulz and G. Taujanskas, \textit{One-dimensional Carrollian fluids~III: global existence and weak continuity in $L^\infty$}, CPHT-RR028.052024,
\href{https://arxiv.org/abs/2407.05972}{arXiv:2407.05972 [math.AP].}


  
  \bibitem{Campoleoni:2018ltl}
A. Campoleoni, L. Ciambelli, C. Marteau, P.M. Petropoulos and K. Siampos \textit{Two-dimensional fluids and their holographic duals}, 
\href{https://www.sciencedirect.com/science/article/pii/S0550321319301786?via\%3Dihub}{Nucl. Phys. \textbf{B946}  (2019) 114692}, \href{https://arxiv.org/abs/1812.04019}{arXiv:1812.04019 [hep-th].}

\bibitem{CMPR}
L.~Ciambelli, C.~Marteau, P.~M. Petropoulos, and R.~Ruzziconi, \textit{Gauges in
  three-dimensional gravity and holographic fluids},   \href{https://link.springer.com/article/10.1007/JHEP11\%282020\%29092}{JHEP {\bf 11} (2020) 092}, \href{http://arxiv.org/abs/2006.10082}{arXiv:2006.10082 [hep-th]}.

\bibitem{Eckart} C. Eckart, \textit{The thermodynamics of irreversible processes III. Relativistic theory of the simple fluid}, \href{https://journals.aps.org/pr/pdf/10.1103/PhysRev.58.919}{Phys. Rev. \textbf{58} (1940) 919.}

\bibitem{Landau}
  L.~D. Landau et E.~M. Lifchitz,
  \textsl{Physique Th\'eorique} Vol. \textbf{6} \textit{M\'ecanique des fluides}, Editions Mir, Moscou, 1969.

 \bibitem{Kovtun:2012rj}
  P.~Kovtun, \textit{Lectures on hydrodynamic fluctuations in relativistic theories}, 
  \href{https://iopscience.iop.org/article/10.1088/1751-8113/45/47/473001}{J. Phys. {\bf A45} (2012) 473001},
\href{http://arxiv.org/abs/1205.5040}{arXiv:1205.5040 [hep-th].}
 
\bibitem{BigFluid}
A. C. Petkou, P. M. Petropoulos, D. Rivera-Betancour and K. Siampos,
  \textit{Relativistic fluids, hydrodynamic frames and their Galilean versus Carrollian avatars},  \href{https://link.springer.com/content/pdf/10.1007/JHEP09(2022)162.pdf}{JHEP {\bf 09} (2022) 162}, \href{http://arxiv.org/abs/2205.09142}{arXiv:2205.09142 [hep-th].}
 
\bibitem{duval1978}
C. Duval and H.~P. K\"unzle, \textit{Dynamics of continua and particles from general covariance of Newtonian gravitation theory,} \href{https://www.sciencedirect.com/science/article/abs/pii/0034487778900630}{Reports on Mathematical Physics \textbf{13} (1978) 351}.
 
 \bibitem{Geracie:2015xfa}
  M.~Geracie, K.~Prabhu and M.~M.~Roberts,
  {\it Fields and fluids on curved non-relativistic spacetimes},
  \href{https://link.springer.com/article/10.1007/JHEP08(2015)042}{JHEP {\bf 1508} (2015) 042},
  \href{https://arxiv.org/abs/1503.02680}{arXiv:1503.02680 [hep-th].}

 \bibitem{Festu}
  G. Festuccia, D. Hansen, J. Hartong and N.~A. Obers,
  \textit{Torsional Newton--Cartan geometry from the N\oe ther procedure,}
  \href{https://journals.aps.org/prd/abstract/10.1103/PhysRevD.94.105023}{Phys.\ Rev.\  {\bf D94} (2016) 105023},  
  \href{https://arxiv.org/abs/1607.01926v2}{arXiv:1607.01926 [hep-th].}
  
 \bibitem{Armas:2019gnb}
  J.~Armas, J.~Hartong, E.~Have, B.~F.~Nielsen and N.~A.~Obers,
  \textit{Newton--Cartan submanifolds and fluid membranes},
  \href{https://journals.aps.org/pre/abstract/10.1103/PhysRevE.101.062803}{Phys. Rev. \textbf{E101} (2020) 062803},
    \href{https://arxiv.org/abs/1912.01613}{arXiv:1912.01613 [hep-th].}

 \bibitem{CMPPS1}
L. Ciambelli, C. Marteau, A. C. Petkou, P.~M. Petropoulos and K. Siampos, \textit{Covariant Galiliean versus Carrollian hydrodynamics from relativistic fluids},    
\href{https://iopscience.iop.org/article/10.1088/1361-6382/aacf1a/meta}{Class.\ Quant.\ Grav.\  {\bf 35} (2018) 165001},
\href{https://arxiv.org/abs/1802.05286}{arXiv:1802.05286 [hep-th].}

 \bibitem{Hartong:2015xda}
  J.~Hartong,
 \textit{Gauging the Carroll algebra and ultra-relativistic gravity,}
  \href{https://link.springer.com/article/10.1007/JHEP08(2015)069}{JHEP {\bf 1508} (2015) 069},
  \href{http://arxiv.org/abs/1505.05011}{arXiv:1505.05011 [hep-th].}

  \bibitem{dutch}
 J.~de Boer, J.~Hartong, N.~A.~Obers, W.~Sybesma and S.~Vandoren,
 {\it Perfect fluids}, \href{https://scipost.org/10.21468/SciPostPhys.5.1.003}{SciPost Phys. \textbf{5}  (2018) 003},
  \href{https://arxiv.org/abs/1710.04708}{arXiv:1710.04708 [hep-th].}

 
  \bibitem{CMPPS2}
L. Ciambelli, C. Marteau, A.~C. Petkou, P.~M. Petropoulos and K. Siampos, \textit{Flat holography and Carrollian fluids},  
\href{https://link.springer.com/article/10.1007\%2FJHEP07\%282018\%29165}{JHEP {\bf 07} (2018) 165}, \href{http://arxiv.org/abs/1802.06809}{arXiv:1802.06809 [hep-th].}

  
  \bibitem{Campoleoni:2023fug}
A.~Campoleoni, A.~Delfante, S.~Pekar, P.~M.~Petropoulos, D.~Rivera-Betancour and M.~Vilatte,
\textit{Flat from anti de Sitter},  \href{https://link.springer.com/article/10.1007/JHEP12(2023)078}{JHEP {\bf 12} (2023) 078}, 
\href{http://arxiv.org/abs/2309.15182}{arXiv:2309.15182 [hep-th].}

\bibitem{carrollianlegends}
J.~de Boer, J.~Hartong, N.~A.~Obers, W.~Sybesma and S.~Vandoren,
\textit{Carroll stories},
\href{https://link.springer.com/article/10.1007/JHEP09(2023)148}{JHEP \textbf{09} (2023) 148},
 \href{http://arxiv.org/abs/2307.06827}{arXiv:2307.06827 [hep-th].}

  

\bibitem{Campoleoni:2022wmf}
A.~Campoleoni, L.~Ciambelli, A.~Delfante, C.~Marteau, P.~M.~Petropoulos and R.~Ruzziconi, \textit{Holographic Lorentz and Carroll frames}, \href{https://link.springer.com/article/10.1007/JHEP12(2022)007}{JHEP \textbf{12} (2022) 007},
 \href{http://arxiv.org/abs/2208.07575}{arXiv:2208.07575 [hep-th].}


 
\end{thebibliography}
\end{document}